\documentclass[prd,showpacs,superscriptaddress,nofootinbib,floatfix,11pt]{revtex4-2}

\usepackage[utf8]{inputenc}
\usepackage{amsmath,amssymb}
\usepackage{slashed}
\usepackage{subfigure}
\usepackage[colorlinks=true,
breaklinks=true,
urlcolor=magenta,
citecolor=blue]{hyperref}
\usepackage[usenames,dvipsnames]{color}
\usepackage{appendix}
\usepackage{braket,bm}
\usepackage{multirow}
\usepackage{enumitem}
\usepackage{color}
\usepackage{cancel}
\usepackage{orcidlink}
\usepackage{mathrsfs}

\usepackage{graphicx}
\usepackage{tikz}
\usepackage[normalem]{ulem}
\usepackage{booktabs}
\usepackage{array}
\usepackage{overpic}
\usepackage{float}
\usepackage{threeparttable}
\allowdisplaybreaks[4]

\usetikzlibrary{calc}
\usetikzlibrary{intersections}
\usetikzlibrary{trees}
\usetikzlibrary{decorations.pathmorphing}
\usetikzlibrary{decorations.markings}
\usetikzlibrary{arrows.meta}
\usetikzlibrary{patterns}
\tikzset{
   global scale/.style={
      scale=#1,
      every node/.append style={scale=#1}},
   photon/.style={decorate, decoration={snake}, draw=red},
   nucleon/.style={draw=black, postaction={decorate},
      decoration={markings,mark=at position .65 with{\arrow[draw=black]{latex}}}},
   pion/.style={draw=blue, postaction={decorate},
      decoration={markings,mark=at position .55 with{\arrow[draw=blue]{}}}},
    nucleonstar/.style={draw=black, postaction={decorate},
      decoration={markings, mark=at position 0.7 with {\arrow[draw=black]{latex}}}},
    }

\newcommand{\itp}{\affiliation{CAS Key Laboratory of Theoretical Physics, Institute of Theoretical Physics,\\ Chinese Academy of Sciences, Beijing 100190, China}}

\newcommand{\ucas}{\affiliation{School of Physical Sciences, University of Chinese Academy of Sciences, Beijing 100049, China}}

\newcommand{\uestc}{\affiliation{School of Physics, University of Electronic Science and Technology of China, Chengdu 611731, China}}

\newcommand{\phcc}{\affiliation{Peng Huanwu Collaborative Center for Research and Education, Beihang University, Beijing 100191, China}}

\newcommand{\scnt}{\affiliation{Southern Center for Nuclear-Science Theory (SCNT), Institute of Modern Physics,\\
Chinese Academy of Sciences, Huizhou 516000, China}}

\newcommand{\md}{\mathrm{d}}

\begin{document}
\title{Photoproduction of the $X(3872)$ beyond vector meson dominance:\\ the open-charm coupled-channel mechanism} 

\author{Xiong-Hui Cao\orcidlink{0000-0003-1365-7178}}\email[Corresponding author: ]{xhcao@itp.ac.cn}
\itp

\author{Meng-Lin Du\orcidlink{0000-0002-7504-3107}}\email{ du.ml@uestc.edu.cn}
\uestc

\author{Feng-Kun Guo\orcidlink{0000-0002-2919-2064}}\email[Corresponding author: ]{fkguo@itp.ac.cn}
\itp\ucas\phcc\scnt

\begin{abstract}

Hidden-charm exotic hadrons will be searched for and investigated at future electron-ion colliders. 
For instance, the $X(3872)$ can be produced through the exclusive process $\gamma p\to X(3872)p$. 
The vector meson dominance model has been commonly employed in estimating the cross sections of such processes. However, the coupled-channel production mechanism through open-charm meson-baryon intermediate states may play a crucial role. 
To assess the significance of such contributions, we estimate the cross section of the $\gamma p\to X(3872)p$ reaction assuming the coupled-channel mechanism. 
For energies near the threshold, the total cross section is predicted to be of tens of nanobarns for $\gamma p\to X(3872)p$, which can be measured at future experimental facilities.
Furthermore, the open-charm coupled-channel mechanism leads to a distinct line shape of the total cross section that can be utilized to reveal the production dynamics. 

\end{abstract}

\maketitle

\newpage

\section{Introduction}

Investigations into the excitation spectra of charmonium(-like) and bottomonium(-like) sectors, particularly above the open-charm and open-bottom thresholds, have unveiled an abundance of novel states. These states elude explanation within the conventional framework of quark-antiquark ($q \bar{q}$) states, see, e.g., Refs.~\cite{Guo:2017jvc,Olsen:2017bmm,Ali:2017jda,Brambilla:2019esw,Chen:2022asf} for recent reviews. Numerous methods have been proposed to elucidate the nature of these exotic states, encompassing a variety of possibilities such as tetraquark configurations based on diquarks, hybrid states with gluonic excitations, and hadronic molecular structures, among others. However, the characteristics of the vast majority of these states remain poorly understood.

Electromagnetic probes such as photons are expected to be essential to gain new insights on the nature of exotic hadrons, and their interactions with hadronic targets lead to the prolific production of resonances. The near-threshold photoproduction of charmonia has attracted considerable attention due to its potential to shed light on a wide array of physical phenomena, e.g., its possible connection to the trace anomaly contribution to the nucleon mass based on the vector meson dominance (VMD) model~\cite{Kharzeev:1994pz, Kharzeev:1998bz}, hidden-charm pentaquarks~\cite{Wang:2015jsa, HillerBlin:2016odx, Strakovsky:2023kqu}, threshold cusp effects~\cite{Du:2020bqj, JointPhysicsAnalysisCenter:2023qgg}, vector quarkonium–nucleon scattering lengths~\cite{Gryniuk:2016mpk, Strakovsky:2019bev}, and so on. Moreover, in contrast to the $\Lambda_b$ decays, the photoduction of the hidden-charm $P_c$ pentaquarks, $\gamma p \to P_c $, was suggested not affected by three-body dynamics from the triangle singularity (TS)~\cite{Mandelstam:1960zz} (also called anomalous threshold) because it is hard to satisfy the on-shell conditions simultaneously, e.g., as discussed in Refs.~\cite{Wang:2015jsa, Cao:2019kst}. Therefore, it has been believed that photo- or leptoproduction reactions serve as reliable benchmarks for unraveling the genuine resonant nature of some exotic states.

The next generation of lepton-hadron facilities including, for example, the Electron-Ion Collider (EIC)~\cite{AbdulKhalek:2021gbh} and the Electron-Ion Collider in China (EicC)~\cite{Anderle:2021wcy} promise to open new possibilities of a hadron spectroscopy program with higher energies than the experiments at the Jefferson Laboratory and high luminosity to study the plethora of the exotic states. In this work, we will investigate the exclusive photoproduction of the $X(3872)$, also known as $\chi_{c1}(3872)$, as a benchmark of the more general exclusive photoproduction of hidden-charm exotic mesons. While cross sections of exclusive reactions are expected to be smaller than the inclusive ones, the cross sections of which have been estimated in Refs.~\cite{Yang:2021jof,Shi:2022ipx}, the constrained kinematics makes the identification of the signal events less ambiguous and can help us clarify their production mechanism. Assessing the exclusive cross section necessitates a comprehensive understanding of both experimental and theoretical aspects. 
So far the exclusive photoproductions of heavy quarkonia and other hidden-charm hadrons were usually accomplished by extending the VMD model~\cite{Sakurai:1960ju,Gell-Mann:1961jim,Kroll:1967it,sakurai1969currents} to charmonia. 
The VMD model posits that the photoproduction can be approximated by replacing the incident photon with the hadron spectral function, generally modeled as a sum of vector meson propagators, with each term multiplied by a coupling related to the vector meson dileptonic width. For a comprehensive review of the VMD model, we refer to Ref.~\cite{OConnell:1995nse} and references therein. 
However, the appropriateness of extending the VMD model to include vector heavy quarkonia in the photoproduction processes has been challenged~\cite{Du:2020bqj,Xu:2021mju}. The heavy quarkonium coupled to the real or spacelike virtual photon must be significantly off-shell. Accurately accounting for such effects is crucial for determining both the magnitude and the momentum-dependence of the photon to vector heavy quarkonium transition~\cite{Xu:2021mju}.
Furthermore, the large virtuality of the $J/\psi$ converted from the real or spacelike photon questions the reliability of the $J/\psi$-nucleon scattering observables, which by definition are on-shell quantities, extracted from the photoproduction measurement under the VMD assumption.

In Ref.~\cite{Du:2020bqj}, it is proposed that the coupled-channel mechanism, in which the open-charm intermediate channels such as the $\Lambda_c\bar D$ and $\Lambda_c \bar D^*$ with thresholds near the $J/\psi p$ one, might dominate the near-threshold photoproduction of the $J/\psi$.
The unique feature of the coupled-channel mechanism is the threshold cusps exactly at the $\Lambda_c \bar D^{(*)}$ thresholds in the energy dependence of the total cross section of the $\gamma p\to J/\psi p$ process.
The updated GlueX data reported recently in Ref.~\cite{GlueX:2023pev} are in line with such expectations; see also the reanalysis using a couplec-channel $K$-matrix formalism by the JPAC group in Ref.~\cite{JointPhysicsAnalysisCenter:2023qgg}.

In this work, we extend the coupled-channel mechanism proposed in Ref.~\cite{Du:2020bqj} to the case of photoproduction of hidden-charm exotic hadrons. 
As a first attempt, we examine the photoproduction of $X(3872)$ on a proton target. The $X(3872)$~\cite{Belle:2003nnu} is by far the most extensively studied exotic charmonium-like state. Its Breit-Wigner (BW) mass coincides with the $D^0 \bar D^{* 0}$ threshold within uncertainties. Its quantum numbers have been determined to be $J^{PC}=1^{++}$~\cite{LHCb:2013kgk}. One prominent feature of the $X(3872)$ is the huge isospin violation in its decays~\cite{ParticleDataGroup:2022pth}. Virtual photoproduction of the $X(3872)$ with a muon beam was recently explored by COMPASS~\cite{COMPASS:2017wql}, however interestingly, the new resonant structure observed, while having a mass consistent with the $X(3872)$, decays into $J/\psi\pi^+\pi^-$ exhibiting a $\pi^+\pi^-$ invariant mass distribution different from that measured by other experiments for the $X(3872)$. Thus, the structure observed by COMPASS should have a negative $C$ parity and is not the $X(3872)$. Nevertheless, COMPASS measured the upper bound for the $X(3872)$ photoproduction cross section at an average $\gamma p$ center-of-mass (c.m.) energy of $\left\langle W_{\gamma p}\right\rangle=13.7~\mathrm{GeV}$ as $\sigma_{\gamma N \rightarrow X(3872) N^{\prime}} \times \mathcal{B}_{X(3872) \rightarrow J / \psi \pi \pi}<2.9$~pb $(\mathrm{CL}=90 \%)$.

We aim to obtain an estimation of the near-threshold $X(3872)p$ photoproduction amplitude beyond the usual VMD model. For this purpose, we evaluate the $\gamma p\to X(3872)p$ scattering amplitude taking into account a few open-charm (one charmed baryon and one anti-charmed meson) channels in a dispersive formalism. 
To ensure that our predictions remain as unbiased as possible regarding the nature of the exotic states, whenever possible our analysis mainly depends on some measured branching fractions and infers other properties from heavy quark spin symmetry (HQSS).
As will be shown, the considered open-charm coupled-channel mechanism leads to falsifiable predictions in the energy dependence of the total cross section of $\gamma p \to X(3872) p$.
Therefore, although the interplay among the various production mechanisms is subtle, future high-precision data will offer the opportunity to discern the underlying dynamics.

The paper is organized as follows. In Sec.~\ref{sec.charmed baryon}, we will provide a concise review of the singly charmed baryon family, with special emphasis on the enigmatic $\Sigma_c(2800)$ and $\Lambda_c(2940)^+$ states. In Sec.~\ref{sec.CC mechanism}, we will select open-charm channels of interest for the problem at hand and construct the corresponding low-energy effective Lagrangians respecting HQSS. As no data are available so far in the threshold region, we will consider several scenarios providing a rough estimate for the coupling constants. In Sec.~\ref{sec.DR+PW}, we present the dispersive representation of the amplitude and the corresponding partial wave projection in detail. In Sec.~\ref{sec.width}, we examine the analytic continuation and investigate the implications of finite width effects on our results. In Sec.~\ref{sec.TCS}, we discuss the results of our calculations in comparison with the VMD prediction in Ref.~\cite{Albaladejo:2020tzt}.  Sec.~\ref{sec.summary} consists of a brief summary. Some technical details on the determination of the coupling constants and the spin-$\frac{3}{2}$ Rarita-Schwinger field are relegated to Appendices~\ref{app.1} and~\ref{app.2}, respectively.

\section{Theoretical framework}

\subsection{Singly charmed baryons}\label{sec.charmed baryon}

Motivated by the evidence for open-charm effects in the $J/\psi$ photoproduction~\cite{Du:2020bqj,GlueX:2023pev,JointPhysicsAnalysisCenter:2023qgg}, we apply the coupled-channel mechanism to the near-threshold $X(3872)$ photoproduction. We first review the singly charmed hadrons relevant to our discussions. 
With the efforts of the CLEO, BaBar, $\mathrm{CDF}$, Belle, and $\mathrm{LHCb}$ Collaborations, the $\Lambda_c(2286)^{+}$, $\Lambda_c(2595)^{+}, \Lambda_c(2625)^{+}$, $\Lambda_c(2860)^{+}, \Lambda_c(2880)^{+}, \Lambda_c(2940)^{+}, \Sigma_c(2455), \Sigma_c(2520)$, and $\Sigma_c(2800)$ have been established (for the observations and theoretical interpretations, see the reviews~\cite{Chen:2016spr,Chen:2022asf}). 

The closeness of the $\Sigma_c(2800)$ and $\Lambda_c(2940)^{+}$ states to the $N D$ and $N D^*$ thresholds, respectively, suggests that they might be the corresponding molecular states.
The $\Lambda_{c}(2940)^{+}$ was reported by the BaBar Collaboration in the invariant mass spectrum of the $D^{0}p$ channel~\cite{BaBar:2006itc} with isospin $0$. Subsequently, the Belle Collaboration confirmed the $\Lambda_{c}(2940)^{+}$ in the $\Lambda_{c}^{+} \pi^{+} \pi^{-}$ final state~\cite{Belle:2006xni}. 
In 2017, the LHCb Collaboration analyzed the amplitude of the decay $\Lambda_{b}^{0} \rightarrow {D}^{0} {p} \pi^{-}$ and found the most likely spin-parity $J^P$ assignment of the $\Lambda_{c}(2940)^{+}$ is $\frac{3}{2}^-$, but the spin ${1\over 2}$ to ${7\over 2}$ possibilities cannot be completely excluded~\cite{LHCb:2017jym}.\footnote{Recently, the $\Lambda_c(2910)^+$ was reported by the Belle Collaboration in the decay process $\bar{B}^0\to\Sigma_c(2455)\pi \bar{p}$~\cite{Belle:2022hnm}. Its mass and width are measured to be $(2913.8\pm5.6\pm3.8)$~MeV and $(51.8\pm20.0\pm18.8)$~MeV, respectively. 
In Ref.~\cite{Zhang:2022pxc}, the authors considered the interplay between the compact $u d c$ core and the $D^* N$ channel in an unquenched framework. They interpreted the recently observed $\Lambda_c (2910)^+$~\cite{Belle:2022hnm} and the $\Lambda_c(2940)^+$ as the conventional $2P$ charmed baryons dressed with the $D^* N$ interacting channel. They also argue that the ${\frac{3}{2}}^{-}$ spin-parity assignment is preferred for $\Lambda_c(2910)^+$, while the $\Lambda_c(2940)^+$ is more likely  a $J^P={\frac{1}{2}}^{-}$ state, in conflict with the preferred ${\frac{3}{2}}^{-}$ assignment by LHCb. However, their results rely on the input bare mass of the $udc$ core and the cutoff from the form factor.}

In 2005, the Belle Collaboration initially reported the $\Sigma_{c}(2800)$ in the $ \Lambda_{c} \pi $ channel~\cite{Belle:2004zjl}, which was confirmed later by the BaBar Collaboration~\cite{BaBar:2008get}. In Ref.~\cite{Sakai:2020psu}, by fitting to the invariant mass spectrum of $D^0 p$ in the decay $\Lambda_b \rightarrow p D^0 \pi^{-}$~\cite{LHCb:2017jym}, the authors extracted the scattering length of the $N D$ system ($D^+ n$ and $D^0 p$) in a coupled-channel nonrelativistic effective field theory (NREFT) framework. They found the absolute value of the scattering length in the isovector channel is very large, and obtained a bound state pole in the isospin limit, which is assigned to the $\Sigma_c(2800)$ with $J^P={\frac{1}{2}}^-$, suggesting an $ND$ molecular nature for this state.

In fact, the interpretations of the $\Lambda_{c}(2940)$ and $\Sigma_{c}(2800)$ are still controversial. The mass of the $\Lambda_{c}(2940)^{+}$ is about 60--100~MeV smaller than the $P$-wave excitation of the $\Lambda_c$ in the the Capstick-Isgur quark model calculation~\cite{Capstick:1986ter}, which is similar to what happened for the $\Lambda(1405)$, $D^*_{s0}(2317)$ and $X(3872)$, while $\Sigma_c(2800)$ was  proposed to be a $P$-wave excitation in, e.g., Refs.~\cite{Chen:2016iyi,Wang:2017kfr}. The $\Lambda_{c}(2940)^{+}$ is only about $6$~MeV below the $D^{*0}p$ threshold, which inspired various $D^\ast N$ molecular interpretations~\cite{He:2006is,He:2010zq,Ortega:2012cx,Dong:2009tg,Dong:2010xv,Zhang:2012jk,Zhao:2016zhf,Wang:2020dhf} within different spin-parity assignments. Similar to the $ \Lambda_{c}(2940)^{+}$, the $\Sigma_c(2800)$ is located just below the $DN$ threshold and was proposed as a candidate of the $DN$ molecule~\cite{Jimenez-Tejero:2009cyn,Zhao:2016zhf,Wang:2020dhf,Dong:2010gu}. It was worth noting that Ref.~\cite{Jimenez-Tejero:2009cyn} used the $t$-channel vector meson exchange model in a coupled-channel approach and concluded that the $\Sigma_c(2800)$ can be interpreted as a dynamically generated resonance with a dominant $D N$ configuration and $J^P={\frac{1}{2}}^{-}$. In the present work, we only focus on the ${\frac{3}{2}}^{-}$ assignment for $\Lambda_c(2940)^+$ and the ${\frac{1}{2}}^{-}$ assignment for $\Sigma_c(2800)$, respectively.

\subsection{Coupled-channel mechanism}\label{sec.CC mechanism}

Firstly, there are six open-charm channels whose thresholds are close to that of $X(3872)p$, i.e., about 4810~MeV:
\begin{align*}
    \Bar{D}^0 \Lambda_c(2940)^+:&\ 4804^{+2}_{-1}~\text{MeV},\\
    \Bar{D}^{*0}\Sigma_c(2800)^+:&\ 4799^{+14}_{-5}~\text{MeV},\\
    D^{*-}\Sigma_c(2800)^{++}:&\ 4811^{+4}_{-6}~\text{MeV},\\
    \Bar{D}^{*0} \Lambda_c(2860)^+:&\ 4863^{+2}_{-6}~\text{MeV},\\
    \Bar{D}^{*0} \Lambda_c(2880)^+:&\ 4888^{+1}_{-0}~\text{MeV},\\
    \Bar{D}^{*0} \Lambda_c(2940)^+:&\ 4946^{+1}_{-1}~\text{MeV},
\end{align*}
where the masses are taken from the Review of Particle Physics (RPP)~\cite{ParticleDataGroup:2022pth}. Due to the sizeable uncertainties from the BW masses and widths of the $\Sigma_c(2800)$ states~\cite{ParticleDataGroup:2022pth}, we choose to take the central value of the pole position and the spin-parity assignment of $\Sigma_c(2800)^+$: $2801.8-i2.6~$MeV\footnote{Actually, Ref.~\cite{Sakai:2020psu} finds two poles, pole-I: $2801.8^{+1.0}_{-4.0}-i(2.6\pm2.6)~$MeV and pole-II: $2807.0^{+4.5}_{-17.1}-i\left(9.4^{+10.2}_{-9.4}\right)~$MeV, in a $pD^0$-$nD^+$ coupled-channel analysis. Both states couple dominantly to the isovector channel, and in the isospin limit only one pole is left. We adopt the central value of pole-I to approximate the isospin partner $\Sigma_c(2800)^+$ and $\Sigma_c(2800)^{++}$. For the $D^{(*)}$ sector, mass differences from isospin breaking are also taken into account, to ensure the feature that one $D^{* 0} \Sigma_c(2800)^+$ threshold ($4809~$MeV) is slightly below $X(3872)p$ and another $D^{* -} \Sigma_c(2800)^{++}$ threshold ($4812$~MeV) is slightly above $X(3872)p$.} and $J^P=\frac{1}{2}^-$ from the coupled-channel analysis based on NREFT~\cite{Sakai:2020psu}. One observes that the small pole width, as reported in \cite{Sakai:2020psu}, is compatible with the predictions from the phenomenological vector meson exchange model outlined in \cite{Jimenez-Tejero:2009cyn}. 
We will not consider the $\Bar{D}^{*0} \Lambda_c(2880)^+$ channel in the following. It is primarily because the $\Lambda_c(2880)^+$ couples to $D^0 p$ in $F$-wave. In fact, we calculated the contribution of the $\Bar{D}^{*0}\Lambda_c(2880)^+$ channel to the cross section and found that it is approximately a few hundred to a thousand times smaller than that of $\Bar{D}^{*0}\Lambda_c(2860)^+$. 

\begin{figure}[tbh]
    \centering
    \includegraphics[width=.5\textwidth]{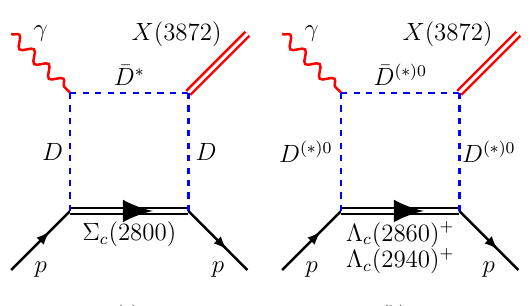} 
    \caption{Feynman diagrams for the proposed coupled-channel mechanism. The left panel illustrates the $\Sigma_c(2800)^+ \bar{D}^{*0}$ and $\Sigma_c(2800)^{++} \bar{D}^{*-}$ intermediate states, while the right panel depicts the $\Lambda_c(2860) \bar{D}^{*0}$ and $\Lambda_c(2940) \bar{D}^{(*)0}$ intermediate states.}\label{fig.Feyn}
\end{figure}
Only the $S$-wave contribution is retained for open-charm systems near the thresholds. We investigate the contribution of the channels shown in Fig.~\ref{fig.Feyn} to the $X(3872)$ photoproduction.
We begin by constructing effective Lagrangians fulfilling invariance under C, P and the isospin symmetry (for strong interaction). In the near-threshold region, it is sufficient to consider Lagrangians that have the lowest number of derivatives, which are given as follows~\cite{He:2011jp,Lin:2021wrb, Meng:2014ota, Du:2020bqj},\footnote{For an easy comparison with the non-relativistic couplings from Ref.~\cite{Guo:2014taa}, we take a convention for $\mathcal{L}_{\gamma D^{(*)} D^*}$ slightly different from that in Ref.~\cite{Du:2020bqj}; for more details see Appendix~\ref{app.1}.}
\begin{align}
    \mathcal{L}_{\gamma D^{(*)} D^*}= & -ig_{\gamma D^0 D^{*0}} F_{\mu \nu} \epsilon^{\mu \nu \alpha \beta}\left(\bar{D}_\alpha^{*0} \overleftrightarrow{\partial_\beta} D^0 +\bar{D}^0 \overleftrightarrow{\partial_\beta} D_\alpha^{*0}\right)-i g_{\gamma D^{*0} D^{*0}} F^{\mu \nu} \bar{D}_\mu^{*0} D_\nu^{*0} \nonumber\\
    & -ig_{\gamma D^+ D^{*-}} F_{\mu \nu} \epsilon^{\mu \nu \alpha \beta}\left(D_\alpha^{*-} \overleftrightarrow{\partial_\beta} D^+ +D^- \overleftrightarrow{\partial_\beta} D_\alpha^{*+}\right),\label{eq.L1}\\
    \mathcal{L}_{Dp\Lambda_{c1}}= &\, ig_{Dp\Lambda_{c1}} \left(\bar{\Lambda}_{c1}^\mu  p\partial_\mu D^0-\bar{p} \Lambda_{c1}^\mu \partial_\mu \bar{D}^0\right)+ig_{D^* p\Lambda_{c1}}\left(\bar{\Lambda}_{c1}^\mu \gamma_5\gamma^\nu p\partial_\mu D^{*0}_\nu-\bar{p} \gamma_5\gamma^\nu \Lambda_{c1}^\mu \partial_\mu \bar{D}^{*0}_\nu\right),\label{eq.L5}\\
    \mathcal{L}_{Dp\Lambda_{c2}}= &\, ig_{Dp\Lambda_{c2}} \left(\bar{\Lambda}_{c2}^\mu\gamma_5  p\partial_\mu D^0+\bar{p}\gamma_5  \Lambda_{c2}^\mu \partial_\mu \bar{D}^0\right) -ig_{D^* p\Lambda_{c2}}\left(\bar{\Lambda}_{c2}^\mu \gamma^\nu p\partial_\mu D^{*0}_\nu-\bar{p} \gamma^\nu \Lambda_{c2}^\mu \partial_\mu \bar{D}^{*0}_\nu\right),\label{eq.L2}\\
    \mathcal{L}_{Dp \Sigma_c}= & -g_{Dp \Sigma_c} \left(\Bar{p}\Sigma_c^+ \Bar{D}^0+\sqrt{2}\Bar{p}\Sigma_c^{++}D^-+\Bar{\Sigma}^+pD^0+\sqrt{2}\Bar{\Sigma}_c^{++}pD^+\right),\label{eq.L3}\\
    \mathcal{L}_{X D \bar{D}^*}= &\, g_{X DD^*} X^\mu\left(\bar{D} D_\mu^*-\bar{D}_\mu^* D\right),\label{eq.L4}
\end{align}
where $A \overleftrightarrow{\partial} B \equiv A(\vec{\partial} B)-(\vec{\partial} A) B$, $N, \Lambda_{c1},\Lambda_{c2}, \Sigma_c^+ (\Sigma_c^{++})$, and $D(D^*)$ are the isodoublet nucleon, the $\Lambda_c(2860)^+$, the $\Lambda_c(2940)^+$, the $\Sigma_c(2800)^+ (\Sigma_c(2800)^{++})$, and the isodoublet $D(D^*)$ meson fields, in order, with the definitions $N=\left(p,n\right)^T$ and $\bar{N}=(\bar{p}, \bar{n})$. The SU(2) vectors $D$ and $D^*$ gather the isospin doublets, $D=\left(-D^{+}, D^0\right)^T, D^*=\left(-D^{*+}, D^*\right)^T$, with the transposed conjugates $\bar{D}=\left(-D^{-} ,\bar{D}^{0}\right), \bar{D}^*=\left(-D^{*-} ,\bar{D}^{* 0}\right)$. 

The Lagrangian $\mathcal{L}_{\gamma D^{(*)} D^*}$ in Eq.~\eqref{eq.L1} contains only magnetic interactions. The coupling $g_{\gamma DD^*}$ can be fixed directly from the data on the experimentally measured total width of the $D^{*+}$ meson (the unknown total width of the $D^{* 0}$ meson is evaluated using isospin symmetry~\cite{Dong:2008gb,Rosner:2013sha,Guo:2019qcn}) and the branching fraction of the decay $D^{* 0} \rightarrow D^0 \gamma$~\cite{ParticleDataGroup:2022pth}. The coupling $g_{\gamma D^* D^*}$ is related to $g_{\gamma D D^*}$ through HQSS~\cite{Amundson:1992yp}. The determination of the couplings is discussed in detail in Appendix~\ref{app.1}.
\begin{table}[tb]
    \centering
    \caption{Values of the effective couplings in the Lagrangians in Eqs.~\eqref{eq.L1}-\eqref{eq.L4} used in the calculation. All parameters are given in appropriate powers of GeV.}
    \begin{ruledtabular}
    \begin{tabular}{llllllllll}
    Coupling & $g_{\gamma D^0 D^{*0}}$ & $g_{\gamma D^{*0} D^{*0}}$ & $g_{\gamma D^{+} D^{*-}}$ & $g_{D^{(*)} p \Lambda_{c1}}$ & $g_{D^{(*)} p \Lambda_{c2}}$ & $g_{DN\Sigma_c}$ & $g_{X D \Bar{D}^*}$ \\
    \hline Value & 0.142 & 0.852 & $-0.035$ & 9.91 & 11.25 
    & 1.57 & 1.98 \\
    Source & \multicolumn{3}{l}{$D^*\to D\gamma$~\cite{ParticleDataGroup:2022pth}, HQSS} & \multicolumn{2}{l}{$\Lambda_{ci}^{+}\to D^0 p$, HQSS} & NREFT~\cite{Sakai:2020psu} & $D^0 \bar{D}^{* 0}$, $J / \Psi \pi^{+} \pi^{-}$ fit~\cite{Meng:2014ota}    \end{tabular}
    \end{ruledtabular}
    \label{tab.coupling}
\end{table}
It is noticed that in the heavy-quark limit, the relation $g_{D N \Lambda_{ci}}=g_{D^* N \Lambda_{ci}}$ ($i=1,2$) holds if $\Lambda_{ci}$ is a conventional baryon.\footnote{Due to the mass of $\Lambda_c(2940)^+$ is slightly below the $D^* p$ threshold, someone considered it to be a standard dynamically generated molecular state~\cite{He:2006is,He:2010zq,Ortega:2012cx,Dong:2009tg,Dong:2010xv,Zhang:2012jk,Zhao:2016zhf,Wang:2020dhf} and has a slightly greater coupling $g_{D^* N \Lambda_{c2}}$ than $g_{D N \Lambda_{c2}}$. However, obtaining the coupling constant $g_{D^* N \Lambda_{c2}}$ through methodologies analogous to those used for $\Sigma_c(2800)$ given in Appendix~\ref{app.1} proves challenging due to the paucity of reliable estimation on the pole residue. Consequently, we resort to HQSS, under the premise that the associated uncertainties will not critically influence the outcomes.} 
It is worth mentioning that the coupling of the $X(3872)$ to $DD^*$ is compatible with the values in Ref.~\cite{Sakai:2020crh} and references therein. 
Since we only aim at an order-of-magnitude estimate of the cross section and its qualitative behavior, we will not consider the errors of these couplings.

\subsection{Dispersion relation and partial-wave projection}\label{sec.DR+PW}

In the following, we use $\Sigma_c,\Lambda_{c1}$, $\Lambda_{c2}$ and $X$ to represent the $\Sigma_c(2800)$, $\Lambda_c(2860)^+$, $\Lambda_c(2940)^+$ and $X(3872)$ states, respectively.
The partial wave (PW) amplitudes for the box diagram contributions in Fig.~\ref{fig.Feyn} are evaluated using the dispersion relation and unitarity relation (for reviews, see, e.g., Refs.~\cite{Oller:2019opk,Yao:2020bxx}) as
\begin{align}\label{eq.int1}
    \mathcal{A}^{(J)}_{\ell S;\Bar{\ell}\Bar{S}~(\gamma p\to Xp)}(s)=\sum_{\ell^\prime,S^\prime}\frac{1}{\pi} \int_{s_\mathrm{th}}^{s_{\mathrm{cut}}} \frac{\mathcal{A}^{(J)}_{\ell^\prime S^\prime;\Bar{\ell}\Bar{S}~(\gamma p\to \bar{D}^* \Sigma_c/\Lambda_c)}\left(s^{\prime}\right) \rho\left(s^{\prime}\right) \mathcal{A}^{(J)}_{\ell^\prime S^\prime;\ell S~(X p\to \bar{D}^* \Sigma_c/\Lambda_c)}\left(s^{\prime}\right)}{s^{\prime}-s} \md s^\prime ,
\end{align}
where the time-reversal symmetry is assumed, e.g., $\mathcal{A}^{(J)}_{\ell^\prime S^\prime;\ell S~(X p\to \bar{D}^* \Sigma_c/\Lambda_c)}=\mathcal{A}^{(J)}_{\ell S;\ell^\prime S^\prime~(\bar{D}^* \Sigma_c/\Lambda_c\to X p)}$, and the tree-level amplitudes are real. Moreover, the thresholds $s_\text{th}=\left(m_{D^{(*)}}+m_{\Sigma_c/\Lambda_c}\right)^2$, and two-body phase space factor $\rho=2 q_\text{cm}/\sqrt{s}$ with $q_\text{cm}$ the magnitude of the three-momentum evaluated in c.m. frame. The dispersive integral in Eq.~\eqref{eq.int1} is evaluated using a hard cutoff at
\begin{align}\label{eq.cutoff}
    \sqrt{s_{\mathrm{cut}}}=\sqrt{q_{\mathrm{max}}^2+m_{\Sigma_c/\Lambda_c}^2}+\sqrt{q_{\max }^2+m_{D^{(*)}}^2},
\end{align}
with a natural value of $q_{\mathrm{max}}$ about $1$~GeV. 

The tree-level amplitudes are given as
\begin{align}
    &\,i\mathcal{A}\left(\gamma(q) p(p_N)\to \Bar{D}^{* 0}(p_D)\Sigma_c(2800)^+(p_\Sigma)\right) \nonumber\\ 
    =&\,4g_{\gamma D^0 D^{*0}} g_{Dp\Sigma_c} \left(\epsilon^{\mu\nu\alpha\beta}\epsilon_\mu\epsilon_\alpha^* \frac{q_\nu p_{D\beta}}{2q\cdot p_D+m_{D^0}^2-m_{D^{*0}}^2}\right)\Bar{u}_\Sigma u_N,\\
    &\,i\mathcal{A}\left(\gamma(q) p(p_N)\to D^{* -}(p_D)\Sigma_c(2800)^{++}(p_\Sigma)\right) \nonumber\\ 
    =&\,4\sqrt{2}g_{\gamma D^+ D^{*-}} g_{Dp\Sigma_c} \left(\epsilon^{\mu\nu\alpha\beta}\epsilon_\mu\epsilon_\alpha^* \frac{q_\nu p_{D\beta}}{2q\cdot p_D+m_{D^+}^2-m_{D^{*-}}^2}\right)\Bar{u}_\Sigma u_N,\\
    &\,i\mathcal{A}\left(\gamma(q) p(p_N)\to \bar{D}^{0}(p_D)\Lambda_c(2940)^{+}(p_\Lambda)\right) \nonumber\\
    =&\,4g_{\gamma D^0 D^{*0}} g_{D^*p\Lambda_{c2}} \epsilon^{\mu\sigma\rho\alpha} \epsilon_\mu \frac{q_\sigma p_{D \rho} (q-p_D)_\nu}{2q\cdot p_D+m_{D^{*0}}^2-m_{D^{0}}^2} \bar{u}_\Lambda^\nu \gamma_\alpha u_N,\\
    &\,i\mathcal{A}\left(\gamma(q) p(p_N)\to \bar{D}^{*0}(p_D)\Lambda_c(2860)^{+}(p_\Lambda)\right) \nonumber\\
    =&\,4 g_{\gamma D^0 D^{*0}}g_{Dp\Lambda_{c1}}\left( \epsilon^{\mu\nu\alpha\beta}\epsilon_\mu\epsilon_\nu^*\frac{q_\alpha p_{D\beta}}{2q\cdot p_D+m_{D^0}^2-m_{D^{*0}}^2}\right)(q-p_D)_\rho\Bar{u}^\rho_\Lambda u_N \nonumber\\
    &+\frac{ig_{\gamma D^{*0}D^{*0}} g_{D^* p \Lambda_{c1}}(q-p_D)_\alpha}{2m_{D^{*0}}^2 q\cdot p_D}\left[m_{D^{*0}}^2\left(\epsilon^*\cdot q \Bar{u}_\Lambda^\alpha \gamma_5\slashed{\epsilon} u_N-\epsilon\cdot\epsilon^* \Bar{u}_\Lambda^\alpha \gamma_5\slashed{q} u_N\right)\right.\nonumber\\
    &\left.-\left(m_{\Lambda_c}+m_N\right)\left(\epsilon\cdot p_D \epsilon^*\cdot q-\epsilon\cdot\epsilon^* q\cdot p_D\right)\Bar{u}_\Lambda^\alpha \gamma_5 u_N\right],\\
    &\,i\mathcal{A}\left(\gamma(q) p(p_N)\to \bar{D}^{*0}(p_D)\Lambda_c(2940)^{+}(p_\Lambda)\right) \nonumber\\
    =&\,4 g_{\gamma D^0 D^{*0}}g_{Dp\Lambda_{c2}}\left( \epsilon^{\mu\nu\alpha\beta}\epsilon_\mu\epsilon_\nu^*\frac{q_\alpha p_{D\beta}}{2q\cdot p_D+m_{D^0}^2-m_{D^{*0}}^2}\right)(q-p_D)_\rho\Bar{u}^\rho_\Lambda\gamma_5 u_N \nonumber\\
    &-\frac{ig_{\gamma D^{*0}D^{*0}} g_{D^* p \Lambda_{c2}}(q-p_D)_\alpha}{2m_{D^{*0}}^2 q\cdot p_D}\left[m_{D^{*0}}^2\left(\epsilon^*\cdot q \Bar{u}_\Lambda^\alpha \slashed{\epsilon} u_N-\epsilon\cdot\epsilon^* \Bar{u}_\Lambda^\alpha \slashed{q} u_N\right)\right.\nonumber\\
    &\left.+\left(m_{\Lambda_c}-m_N\right)\left(\epsilon\cdot p_D \epsilon^*\cdot q-\epsilon\cdot\epsilon^* q\cdot p_D\right)\Bar{u}_\Lambda^\alpha u_N\right],\\
    &\,i\mathcal{A}\left(X(3872)(p_X) p(p_N)\to \bar{D}^{*0}(p_D)\Sigma_c(2800)^{+}(p_\Lambda)\right) = -ig_{X DD^*} g_{Dp\Sigma_c}  \frac{\epsilon\cdot \epsilon^*\,\Bar{u}_\Sigma u_N}{(p_X-p_D)^2-m^2_{D^0}},\\
    &\,i\mathcal{A}\left(X(3872)(p_X) p(p_N)\to D^{*-}(p_D)\Sigma_c(2800)^{++}(p_\Lambda)\right) = -i\sqrt{2}g_{X DD^*} g_{Dp\Sigma_c} \frac{\epsilon\cdot \epsilon^* \, \Bar{u}_\Sigma u_N}{(p_X-p_D)^2-m^2_{D^0}},\\
    &\,i\mathcal{A}\left(X(3872)(p_X) p(p_N)\to \bar{D}^{0}(p_D)\Lambda_c(2940)^{+}(p_\Lambda)\right) \nonumber\\
    =&-\frac{ig_{X DD^*} g_{D^* p\Lambda_{c2}}(p_X-p_D)_\alpha}{m_{D^{*0}}^2 \left((p_X-p_D)^2-m^2_{D^{*0}}\right)}\left[m_{D^{*0}}^2\Bar{u}_\Lambda^\alpha \slashed{\epsilon} u_N+\epsilon\cdot p_D\left(m_{\Lambda_c}-m_N\right)\Bar{u}_\Lambda^\alpha u_N\right],\\
    &\,i\mathcal{A}\left(X(3872)(p_X) p(p_N)\to \bar{D}^{*0}(p_D)\Lambda_c(2860)^{+}(p_\Lambda)\right) = ig_{X DD^*}g_{Dp\Lambda_{c1}}\frac{\epsilon\cdot\epsilon^*\, (p_X-p_D)_\mu}{(p_X-p_D)^2-m^2_{D^0}}\Bar{u}_\Lambda^\mu u_N,\\
    &\,i\mathcal{A}\left(X(3872)(p_X) p(p_N)\to \bar{D}^{*0}(p_D)\Lambda_c(2940)^{+}(p_\Lambda)\right) = ig_{X DD^*}g_{Dp\Lambda_{c2}}\frac{\epsilon\cdot\epsilon^*\, (p_X-p_D)_\mu}{(p_X-p_D)^2-m^2_{D^0}}\Bar{u}_\Lambda^\mu\gamma_5 u_N,
\end{align}
where the amplitudes with photon are manifestly gauge invariant, and the vector propagator \begin{align}
    G_{D^{* 0}}\left(q^2\right)=\frac{i\left(-g^{\mu \nu}+\frac{q^\mu q^\nu}{m_{D^{* 0}}^2}\right)}{q^2-m_{D^{* 0}}^2}
\end{align} 
is used. Other conventions especially for the Rarita-Schwinger vector-spinors $u^\alpha_\Lambda$ are listed in Appendix~\ref{app.2}.

To take into account the potentially large off-shellness of the exchanged particles, we augment them with a monopole form factor~\cite{Colangelo:2003sa,Cheng:2004ru,Du:2020bqj}
\begin{align}\label{eq.FF}
    F(t)=\frac{\Lambda^2-m_{\mathrm{ex}}^2}{\Lambda^2-t},
\end{align}
with $m_{\mathrm{ex}}$ the mass of the exchanged particle. A natural value for the cutoff $\Lambda$ is the mass of the lowest neglected exchange particle, so that we set~\cite{Cheng:2004ru} $\Lambda=m_{\mathrm{ex}}+\eta \Lambda_{\mathrm{QCD}}$ with $\Lambda_{\mathrm{QCD}} \simeq 250~\mathrm{MeV}$ and the parameter $\eta$ which depends on both exchanged and external particles~\cite{Cheng:2004ru} is expected to be of order unity. For simplicity, if not stated otherwise, we set $\eta=1$ and $\Lambda_{\mathrm{QCD}}=250 ~\mathrm{MeV}$ for exchanged particle $D^{(*)}$.

We utilize the covariant $J\ell S$ PW scheme from Refs.~\cite{Gulmez:2016scm, Oller:2019rej}, the general PW projection for the process $\bar{1}+\Bar{2}\to 1+2$ in the $J\ell S$ basis is given by\footnote{We use the following normalization of the PW amplitudes, $\operatorname{Im} T_{\ell S ; \bar{\ell} \bar{S}}^{(J I)}=\sum_{\ell^{\prime \prime}, S^{\prime \prime}} \frac{2\left|\boldsymbol{p}^{\prime \prime}\right|}{\sqrt{s}} T_{\ell, S ; \ell^{\prime \prime}, S^{\prime \prime}}^{(J I)} T_{\ell^{\prime \prime}, S^{\prime \prime} ; \bar{\ell} \bar{S}}^{(J I)^*}$, which is different from the one used in Refs.~\cite{Gulmez:2016scm, Oller:2019rej}, $\operatorname{Im} T_{\ell S ; \bar{\ell} \bar{S}}^{(J I)}=-\sum_{\ell^{\prime \prime}, S^{\prime \prime}} \frac{\left|\boldsymbol{p}^{\prime \prime}\right|}{8 \pi \sqrt{s}} T_{\ell, S ; \ell^{\prime \prime}, S^{\prime \prime}}^{(J I)} T_{\ell^{\prime \prime}, S^{\prime \prime} ; \bar{\ell} \bar{S}}^{(J I)^*}$.}
\begin{align}\label{eq.PW1}
    \mathcal{A}_{\ell S ; \bar{\ell} \bar{S}}^{(J)}(s)= & \frac{Y_{\bar{\ell}}^0\left(\hat{\boldsymbol{z}}\right)}{16\pi(2 J+1)} \sum_{\sigma_1, \sigma_2, \bar{\sigma}_1, \bar{\sigma}_2, m} \int \md \hat{\boldsymbol{p}}~Y_{\ell}^m(\hat{\boldsymbol{p}})^*\left(\sigma_1 \sigma_2 M \mid s_1 s_2 S\right)(m M \bar{M} \mid \ell S J)\left(\bar{\sigma}_1 \bar{\sigma}_2 \bar{M} \mid \bar{s}_1 \bar{s}_2 \bar{S}\right) \nonumber\\
    & \times(0 \bar{M} \bar{M} \mid \bar{\ell} \bar{S} J) \mathcal{A}\left(p_1, p_2, \bar{p}_1, \bar{p}_2, \sigma_1, \sigma_2, \bar{\sigma}_1, \bar{\sigma}_2\right),
\end{align}
with the total angular momentum $J$, the orbital angular momenta $\Bar{\ell},\ell$, and the total spins $\Bar{S},S$, where only $J$ is a good quantum number. $\Bar{M}$ $(M)$ and  $\Bar{\sigma}_i$ ($\sigma_i$) correspond to the third components of $\bar{S}$ ($S$) and $\bar{s}_i$ $(s_i)$, respectively, with $\Bar{M}=\Bar{\sigma}_1+\Bar{\sigma}_2$ ($ M=\sigma_1+\sigma_2$). The Clebsch-Gordan (CG) coefficient $\left(m_1 m_2 m_3 \mid j_1 j_2 j_3\right)\equiv \left\langle j_1,m_1 ; j_2, m_2 \mid j_3, m_3\right\rangle$ refers to the composition $\boldsymbol{j}_1+\boldsymbol{j}_2=\boldsymbol{j}_3$ with $m_i$ the third component of $\boldsymbol{j}_i$. 
The Mandelstam variables are defined by 
\begin{align}
    s=(\Bar{p}_1+\Bar{p}_2)^2,\quad t=(\Bar{p}_1-p_1)^2, \quad u=(\Bar{p}_1-p_2)^2.
\end{align}
for the scattering of the $\Bar{1} (\gamma/X(\Bar{p}_1, \Bar{s}_1))+\Bar{2} (p(\Bar{p}_2, \Bar{s}_2))\to 1 (\Bar{D}^*(p_1, s_1))+2 (\Sigma_c/\Lambda_{c}(p_2, s_2))$ processes.\footnote{The discussions of the process $\Bar{1} (\gamma/X(\Bar{p}_1, \Bar{s}_1))+\Bar{2} (p(\Bar{p}_2, \Bar{s}_2))\to 1 (\Bar{D}(p_1, s_1))+2 (\Lambda_{c2}(p_2, s_2))$ are analogous.}
The calculation is carried out in the c.m. frame. 
We choose the initial three-momentum in the c.m. frame to be along the $z$-axis, and the final-state momentum in the $xOz$ plane, at a polar angle $\theta$ with respect to the $z$-axis,
\begin{align}
    \begin{aligned}
    & \Bar{p}_1^\mu=\left(\Bar{E}_1, 0,0, |\bar{\boldsymbol{p}}|\right), \quad &&\Bar{p}_2^\mu=\left(\Bar{E_2}, 0,0,-|\bar{\boldsymbol{p}}|\right), \\
    & p_1^{\mu}=\left(E_1, |\boldsymbol{p}| \sin \theta, 0, |\boldsymbol{p}| \cos \theta\right), \quad &&p_2^{\mu}=\left(E_2,-|\boldsymbol{p}| \sin \theta, 0,-|\boldsymbol{p}| \cos \theta\right),
    \end{aligned}
\end{align}
where
\begin{gather*}
    \Bar{E}_1=\frac{s+m^2_{\gamma/X}-m_N^2}{2\sqrt{s}},\quad \Bar{E}_2=\frac{s+m_N^2-m^2_{\gamma/X}}{2\sqrt{s}},\\
    E_1=\frac{s+m_{D^*}^2-m^2_{\Sigma_c/\Lambda_c}}{2\sqrt{s}},\quad E_2=\frac{s+m^2_{\Sigma_c/\Lambda_c}-m_{D^*}^2}{2\sqrt{s}},\\
    |\Bar{\boldsymbol{p}}|=\sqrt{\left(\frac{s+m^2_{\gamma/X}-m_N^2}{2\sqrt{s}}\right)^2-m_{\gamma/X}^2},\quad |\boldsymbol{p}|=\sqrt{\left(\frac{s+m_{D^*}^2-m^2_{\Sigma_c/\Lambda_c}}{2\sqrt{s}}\right)^2-m_{D^*}^2}.
\end{gather*}

While only the $S$-wave, i.e., $\ell=0$, is considered for the open-charm systems $\bar{D}^*\Sigma_c(\Lambda_c)$, all $S$-, $P$- and $D$-waves are taken into account for the $\gamma p$ and $Xp$ systems ($\Bar{\ell}=0,1,2$). The PW projection in Eq.~\eqref{eq.PW1} can be simplified as
\begin{align}\label{eq.PW2}
    \mathcal{A}_{0 S ; \bar{\ell} \bar{S}}^{(S)}(s)= &\, \frac{\sqrt{2\Bar{\ell}+1}}{32\pi(2 J+1)} \sum_{\sigma_1, \sigma_2, \bar{\sigma}_1, \bar{\sigma}_2} \int^1_{-1} \md \cos\theta~\left(\sigma_1 \sigma_2 M \mid 1 \frac{1}{2}\left(\frac{3}{2}\right) S\right)\left(\bar{\sigma}_1 \bar{\sigma}_2 M \mid 1 \frac{1}{2} \bar{S}\right) (0 M M \mid \bar{\ell} \bar{S} S)\nonumber\\
    & \times \mathcal{A}\left(p_1(s,\cos\theta), p_2(s,\cos\theta), \bar{p}_1(s), \bar{p}_2(s), \sigma_1, \sigma_2, \bar{\sigma}_1, \bar{\sigma}_2\right).
\end{align}

In the following, we detail the polarization vectors of the particles mentioned above. The polarizations of the massless photon $\gamma$ take the form 
\begin{align}
    \epsilon^\mu(\Bar{p}_1,\Bar{\sigma}_1=1)=-\frac{1}{\sqrt{2}}(0, 1, i, 0), \quad \epsilon^\mu(\Bar{p}_1,\Bar{\sigma}_1=-1)=\frac{1}{\sqrt{2}}(0, 1,-i, 0).
\end{align}
The polarization vectors of the spin-1 meson $X(3872)$ with a mass $m_{X}$ are given by~\cite{Gulmez:2016scm}
\begin{align}
    \epsilon^\mu(\Bar{p}_1,\Bar{\sigma}_1=0)&=\frac{1}{m_{X}}\left(|\bar{\boldsymbol{p}}|, 0, 0, \bar{E}_1\right),\nonumber\\
    \epsilon^\mu(\Bar{p}_1,\Bar{\sigma}_1=1)&=-\frac{1}{\sqrt{2}}(0, 1, i, 0), \quad \epsilon^\mu(\Bar{p}_1,\Bar{\sigma}_1=-1)=\frac{1}{\sqrt{2}}(0, 1,-i, 0).
\end{align}
Accordingly, the polarization vectors of the $\Bar{D}^*$ with a mass $m_{D^*}$ should be taken as~\cite{Gulmez:2016scm}
\begin{align}
    \epsilon^\mu(p_1,\sigma_1=0)&=\left(\frac{|\boldsymbol{p}|}{m_{D^*}} \cos\theta, \frac{1}{2}\left(\frac{E_1}{m_{D^*}}-1\right) \sin 2 \theta, 0, \frac{E_1}{m_{D^*}}\cos^2\theta+\sin^2\theta\right),\nonumber\\
    \epsilon^\mu(p_1,\sigma_1=1)&=\left(-\frac{1}{\sqrt{2}} \frac{|\boldsymbol{p}|}{m_{D^*}} \sin \theta, -\frac{1}{\sqrt{2}}\left(\frac{E_1}{m_{D^*}} \sin ^2 \theta+\cos ^2 \theta\right), -\frac{i}{\sqrt{2}}, -\frac{1}{2 \sqrt{2}}\left(\frac{E_1}{m_{D^*}}-1\right) \sin 2 \theta\right),\nonumber\\
    \epsilon^\mu(p_1,\sigma_1=-1)&=\left(\frac{1}{\sqrt{2}} \frac{|\boldsymbol{p}|}{m_{D^*}} \sin \theta, \frac{1}{\sqrt{2}}\left(\frac{E_1}{m_{D^*}} \sin ^2 \theta+\cos ^2 \theta\right), -\frac{i}{\sqrt{2}}, \frac{1}{2 \sqrt{2}}\left(\frac{E_1}{m_{D^*}}-1\right) \sin 2 \theta\right).
\end{align}
For the spinor of the initial proton, it takes the form of
\begin{align}
    &u(\Bar{p}_2,\Bar{\sigma}_2=\frac{1}{2})=\sqrt{\Bar{E}_2+m_N}\left(1,0,-\frac{|\Bar{\boldsymbol{p}}|}{\Bar{E}_2+m_N},0\right)^T,\nonumber\\
    &u(\Bar{p}_2,\Bar{\sigma}_2=-\frac{1}{2})=\sqrt{\Bar{E}_2+m_N}\left(0,1,0,\frac{|\Bar{\boldsymbol{p}}|}{\Bar{E}_2+m_N}\right)^T.
\end{align}
Similarly, the Dirac spinors of $\Sigma_c/\Lambda_c$ read as
\begin{align}\label{eq.spinor_sigma}
    u(p_2,\sigma_2=\frac{1}{2})&=\sqrt{E_2+m_{\Sigma_c/\Lambda_c}}\left(1,0,-\frac{|\boldsymbol{p}|\cos\theta}{E_2+m_{\Sigma_c/\Lambda_c}},-\frac{|\boldsymbol{p}|\sin\theta}{E_2+m_{\Sigma_c/\Lambda_c}}\right)^T,\nonumber\\
    u(p_2,\sigma_2=-\frac{1}{2})&=\sqrt{E_2+m_{\Sigma_c/\Lambda_c}}\left(0,1,-\frac{|\boldsymbol{p}|\sin\theta}{E_2+m_{\Sigma_c/\Lambda_c}},\frac{|\boldsymbol{p}|\cos\theta}{E_2+m_{\Sigma_c/\Lambda_c}}\right)^T.
\end{align}
Following Eq.~\eqref{eq.vecpor}, the polarization vectors satisfy the transformation property:
\begin{align}
    \epsilon(-\boldsymbol{p},\sigma)=\left(-\epsilon(\boldsymbol{p},\sigma)_0, \boldsymbol{\epsilon}(\boldsymbol{p},\sigma)\right),
\end{align}
and thus the polarization vectors $\epsilon^\mu(p_2,\lambda)$ are given by
\begin{align}
     \epsilon^\mu(p_2,\lambda=0)&=\left(-\frac{|\boldsymbol{p}|}{m_{\Lambda_c}} \cos\theta, \frac{1}{2}\left(\frac{E_2}{m_{\Lambda_c}}-1\right) \sin 2 \theta, 0,  \frac{E_2}{m_{\Lambda_c}}\cos^2\theta+\sin^2\theta\right),\nonumber\\
    \epsilon^\mu(p_2,\lambda=1)&=\left(\frac{1}{\sqrt{2}} \frac{|\boldsymbol{p}|}{m_{\Lambda_c}} \sin \theta, -\frac{1}{\sqrt{2}}\left(\frac{E_2}{m_{\Lambda_c}} \sin ^2 \theta+\cos ^2 \theta\right), -\frac{i}{\sqrt{2}}, -\frac{1}{2 \sqrt{2}}\left(\frac{E_2}{m_{\Lambda_c}}-1\right) \sin 2 \theta\right),\nonumber\\
    \epsilon^\mu(p_2,\lambda=-1)&=\left(-\frac{1}{\sqrt{2}} \frac{|\boldsymbol{p}|}{m_{\Lambda_c}} \sin \theta, \frac{1}{\sqrt{2}}\left(\frac{E_2}{m_{\Lambda_c}} \sin ^2 \theta+\cos ^2 \theta\right), -\frac{i}{\sqrt{2}}, \frac{1}{2 \sqrt{2}}\left(\frac{E_2}{m_{\Lambda_c}}-1\right) \sin 2 \theta\right).
\end{align}
The spin-$\frac{3}{2}$ Rarita-Schwinger vector-spinors can be constructed based on $\epsilon^\mu(p_2,\lambda)$ and $u(p_2,\sigma)$ as~\cite{Huang:2003ym}
\begin{align}
    u^\mu(p_2, \sigma_2)=\sum_{\lambda, \sigma}\left\langle 1, \lambda; \frac{1}{2}, \sigma \right|\left. \frac{3}{2}, \lambda+\sigma\right\rangle \epsilon^\mu(p_2, \lambda) u(p_2, \sigma).
\end{align}

\subsection{Analytic continuation and width impacts}\label{sec.width}

We note that Eq.~\eqref{eq.int1} is a dispersive integral starting from the open-charm threshold and the $\bar{D} \Lambda_c(2940)^+$ and $\bar{D}^* \Sigma_c(2800)^+$ thresholds are slightly below the threshold of $X(3872) p$. Taking $\bar{D}^* \Sigma_c(2800)^+$ as an example, the analytic continuation for amplitudes $\mathcal{A}_{\ell^{\prime} S^{\prime}; \ell S}^{(J)}\left(X p \rightarrow \bar{D}^{* 0} \Sigma_c^+ \right)$ between the two thresholds is subtle. To illustrate this point, we consider the $X p \rightarrow \bar{D}^{* 0} \Sigma_c^+$ scattering through the exchange of a $D^0$ assuming all the particles are stable and neglecting their spins. All the interesting features remain while avoiding the unnecessary kinematical complexity, e.g., from polarizations. As an illustration, we consider only the $S$-wave amplitudes:
\begin{align}\label{eq.A0}
    \mathcal{A}^{(0)}\propto \frac1{Q(s)}\left[\ln\left(P_2(s)+Q(s)\right)-\ln\left(P_2(s)-Q(s)\right)\right],
\end{align}
where
\begin{align}
    P_2(s)&=s^2+\left(2 m_D^2-m_{D^*}^2+m_{X}^2+m_N^2+m_{\Sigma_c}^2\right)s-\left(m_{D^*}^2-m_{X}^2\right)\left(m_N^2-m_{\Sigma_c}^2\right),\\
    Q(s)&=\sqrt{\lambda\left(s,m_{X}^2,m_N^2\right)}\sqrt{\lambda\left(s,m_{D^*}^2,m_{\Sigma_c}^2\right)},
\end{align}
with $\lambda(x, y, z) \equiv x^2+y^2+z^2-2xy-2xz-2yz$. The physical left-hand cuts (LHCs) here stem from the exchange of a $t$-channel $D^0$ (for a more comprehensive discussion of the analytic singularities of PW amplitudes, see Ref.~\cite{Kennedy:1962ovz}), and there are two corresponding logarithmic branch points $s_{\pm}$:
\begin{align}\label{eq.s+-}
    s_\pm=&\,\frac{1}{2m_D^2}\bigg[ m_D^2\left(m_{D^*}^2+m_{X}^2+m_N^2+m_{\Sigma_c}^2\right)+\left(m_{D^*}^2-m_{X}^2\right)\left(m_N^2-m_{\Sigma_c}^2\right)-m_D^4\nonumber\\
    &\left.\pm\sqrt{\lambda\left(m_D^2,m_{N}^2,m_{\Sigma_c}^2\right)}\sqrt{\lambda\left(m_D^2,m_{D^*}^2,m_{X}^2\right)}\right].
\end{align}
Since $m_{\Sigma_c} < m_D + m_N$ and $m_{X} < m_D + m_{D^*}$ (with the central values of the involved particle masses listed in RPP~\cite{ParticleDataGroup:2022pth}), both branch points $s_{\pm}$ are located on the real axis and slightly below the $D^* \Sigma_c$ threshold. Therefore, it seems that $\mathcal{A}^{(0)}$ can be analytically continued to the open-charm $D^* \Sigma_c$ threshold safely. 

We however run immediately into the following difficulty: the expression $\mathcal{A}^{(0)}$ in Eq.~\eqref{eq.A0} has an unphysical singularity corresponding to a threshold divergence. For definiteness, we show that if the mass of the exchanged $D^0$ satisfies the relation,
\begin{align}
    &\sqrt{\frac{m_N m_{D^*}^2+m_{X} m_{\Sigma_c}^2}{m_N+m_{X}}-m_{N}m_{X}}(= 1863.26~\mathrm{MeV})<m_D(=1864.84~\mathrm{MeV})<\nonumber\\
    &\sqrt{\frac{m_{D^*}m_N^2+m_{\Sigma_c} m_{X}^2}{m_{D^*}+m_{\Sigma_c}}-m_{D^*}m_{\Sigma_c}}(= 1865.07~\mathrm{MeV}),
\end{align}
one of the roots $s_P$ from the second-order polynomial $P_2(s)$ will be located between the two thresholds, $(m_{D^*}+m_{\Sigma_c})^2$ and $(m_N+m_{X})^2$, which indicate that $P_2(s)<0$ for $s \in ((m_{D^*}+m_{\Sigma_c})^2,s_P)$ and $P_2(s)>0$ for $s \in (s_P,(m_N+m_{X})^2)$. 
One can immediately find that $Q(s)$ in Eq.~\eqref{eq.A0} is purely imaginary and tends to zero as it approaches either threshold. From the properties of $P_2(s)$ and $Q(s)$, when $s$ approaches $(m_N+m_{X})^2$ from below, one has
\begin{align}
    \lim_{s\to (m_N+m_{X})^2} \mathcal{A}^{(0)}\propto\lim_{\varepsilon\to 0}\frac{\ln\left(P_2((m_N+m_{X})^2\right)+i\varepsilon)-\ln\left(P_2((m_N+m_{X})^2\right)-i\varepsilon)}{\varepsilon}=\text{finite},
\end{align}
where $\varepsilon$ is an infinitesimal positive number. However, when $s$ approaches the lower threshold from above, one has
\begin{align}
    \lim_{s\to (m_{D^*}+m_{\Sigma_c})^2} \mathcal{A}^{(0)}&\propto\lim_{\varepsilon\to 0}\frac{\ln\left(P_2((m_{D^*}+m_{\Sigma_c})^2\right)+i\varepsilon)-\ln\left(P_2((m_{D^*}+m_{\Sigma_c})^2\right)-i\varepsilon)}{\varepsilon}\nonumber\\
    &=\lim_{\varepsilon\to 0}\frac{i\pi-(-i\pi)}{\varepsilon}\to\infty,
\end{align}
which implies that the tree-level amplitude $\mathcal{A}^{(0)}$ has an unphysical threshold divergence.
Higher PWs and the scattering of particles with spins also exhibit this feature. 

Although such a threshold divergence cannot be removed from any redefinition of the logarithmic branch cut, there is no need for concern regarding its influence on the physical region, even without considering the width of any particles. 
This is because the experimental observables are only related to the physical region $((m_N+m_{X})^2,\infty)$ of the $X p \rightarrow \bar{D}^{* 0} \Sigma_c^+$ scattering, which is always above the unphysical singularity.

For an investigation of the energy dependence of the total cross section, it is essential to account for the widths of intermediate states as they can smear the involved threshold cusps, which are square-root branch points, and the triangle singularities, which are subleading Landau singularities of the box diagrams in Fig.~\ref{fig.Feyn} and are logarithmic branch points.
The tiny $D^{*0}$ width about 55.3~keV~\cite{Rosner:2013sha,Guo:2019qcn} can be safely neglected; in contrast, the physical $\Sigma_c(2800)$ has a width of 5.2~MeV from the analysis in Ref.~\cite{Sakai:2020psu}.
A relatively rigorous formalism should take into account three-body intermediate state effects ($DD^* N$). However, Aitchison demonstrated in Ref.~\cite{Aitchison:1964rwb} that, although the singularities of two-body and three-body amplitudes are in general different, the two-body approximation with a complex resonance pole mass is appropriate for calculating the ``enhancement" effects due to singularities of the three-body amplitude, near the physical region. 
Meanwhile, the width should not be excessively large to destroy the continuity in the physical region, ensuring that any LHCs cannot cross the physical region. In practice, the complex mass method is successful for above mentioned intermediate states. It may, however, lead to a sizeable correction for the case of $\bar{D}^{*0}\Lambda_c(2940)^+$ channel. 
An alternative method to take into account this effect relies on convolution with a BW distribution (Lorentzian mass squared distribution)~\cite{Oller:1997ti} for the $\Lambda_c(2940)$. The resulting imaginary part of amplitude denoted by $\operatorname{Im}\tilde{\mathcal{A}}_{\left(\gamma p \rightarrow X p\right)}(s,m_{\Lambda_c}^2)$ (the PW indices are omitted for simplicity) is given by~\cite{Molina:2008jw}\footnote{An alternative expression of $\sigma(m^2,m_{\Lambda_c}^2)$ is also utilized in literature, e.g. Ref.~\cite{Nagahiro:2008mn}. In the narrow-width limit, the spectral function converges to a Dirac $\delta$ distribution, rendering the numerical differences between the different forms negligible within the uncertainties of our results.}
\begin{align}\label{eq.CV}
    \operatorname{Im}\tilde{\mathcal{A}}_{\left(\gamma p \rightarrow X p\right)}(s,m_{\Lambda_{c2}}^2)=\frac{1}{\mathcal{N}}\int^{(m_{\Lambda_{c2}}+2 \Gamma_{\Lambda_{c2}})^2}_{(m_{\Lambda_{c2}}-2 \Gamma_{\Lambda_{c2}})^2}\md m^2~\sigma(m_{\Lambda_{c2}}^2,m^2)\operatorname{Im}\mathcal{A}_{\left(\gamma p \rightarrow X p\right)}(s,m^2)\theta(s-(m+m_{D^*})^2),
\end{align}
where the BW spectral function and the normalization factor are
\begin{align}
    \sigma(s,m_{\Lambda_{c2}}^2)&=-\frac{1}{\pi} \operatorname{Im}\left(\frac{1}{s-m_{\Lambda_{c2}}^2+i m_{\Lambda_{c2}} \Gamma_{\Lambda_{c2}}}\right)=\frac{1}{\pi}\frac{m_{\Lambda_{c2}} \Gamma_{\Lambda_{c2}}}{\left(s-m_{\Lambda_{c2}}^2\right)^2+m_{\Lambda_{c2}}^2 \Gamma_{\Lambda_{c2}}^2},\\
    \mathcal{N}&=\int^{(m_{\Lambda_{c2}}+2 \Gamma_{\Lambda_{c2}})^2}_{(m_{\Lambda_{c2}}-2 \Gamma_{\Lambda_{c2}})^2}\md m^2~\sigma(m_{\Lambda_{c2}}^2,m^2),
\end{align}
with $\Gamma_{\Lambda_{c2}}$ the BW width.

Although the masses of the $X(3872), D^0$ and $D^{* 0}$ are quite well determined by experiments, we still do not know whether the $X(3872)$ mass is above or below the $D^0 \bar{D}^{* 0}$ threshold. With the values in RPP~\cite{ParticleDataGroup:2022pth}, one gets the binding energy
\begin{align}
    \delta \equiv m_{D^0}+m_{D^{* 0}}-m_{X}=(0.04 \pm 0.09)~\mathrm{MeV}.
\end{align}
With the central value of the binding energy is positive, the subprocesses $Xp\to\Bar{D}^{*0} \Sigma_c^+/\Lambda_c^+$ in Eq.~\eqref{eq.int1} cannot exactly exchange an on-shell $t$-channel $D^0$. The LHCs from $s_\pm$ in Eq.~\eqref{eq.s+-} do not overlap with the unitary cut. However, for the case of a negative $\delta$, the two branch points $s_\pm$ are located on the unitary cut, and the former dispersive formalism would not hold without further modifications. Nevertheless, that the branch points $s_\pm$ are present in the physical region implies that the $X(3872)$ meson is unstable and it is necessary to take the finite-width effects into account.
In order to consistently implement the unstable $X(3872)$ effect, one can use the complex mass scheme, which will shift $s_{\pm}$ from the physical cut by a small distance. 
Since the width $\Gamma_{X}$ is small,  the physical amplitudes can still feel the influence of these singularities. 
Although the width of $X(3872)$ given in RPP is $\Gamma_{X}=(1.19\pm 0.21)$~MeV~\cite{ParticleDataGroup:2022pth}, the value is from averaging the BW width parameters from the $J/\psi \pi^+\pi^-$ mode. However, for cases like the one at hand, the $X(3872)$ is located on top of the $D^0\bar D^{*0}$ threshold and couples strongly to this channel in the $S$-wave, the BW parametrization should never be used as it violates unitarity by completely neglecting the probability attributed to this $S$-wave channel; correspondingly, the BW width parameter does not have much physical meaning.
Notice that in the recent BESIII coupled-channel analysis~\cite{BESIII:2023hml}, the half-maximum width of the $X(3872)$ line shape is of the order of $100$~keV, which is compatible with the molecular state model prediction~\cite{Fleming:2007rp,Guo:2014hqa}.

\section{Total cross section and triangle singularity}\label{sec.TCS}

In our chosen normalization, the total cross section is given by
\begin{align}\label{eq.cross section}
    \sigma_{\gamma p\to Xp}(s)&=\frac{16\pi}{2(2 \Bar{s}_2+1)s}\frac{|\boldsymbol{p}(s)|}{|\Bar{\boldsymbol{p}}(s)|}\sum_{J,\ell S,\Bar{\ell}\Bar{S}}(2J+1)\left|\mathcal{A}^{(J)}_{\ell S;\Bar{\ell}\Bar{S}~(\gamma p\to Xp)}(s)\right|^2 \nonumber\\
    &=\frac{4\pi}{s}\frac{|\boldsymbol{p}(s)|}{|\Bar{\boldsymbol{p}}(s)|}\sum_{J,\ell S,\Bar{\ell}\Bar{S}}(2J+1)\left|\mathcal{A}^{(J)}_{\ell S;\Bar{\ell}\Bar{S}~(\gamma p\to Xp)}(s)\right|^2,
\end{align}
where the first factor $1/2$ accounts for the polarizations of the real photon. The c.m. energy is $\sqrt{s}=W_{\gamma p}$, and
\begin{align}
    |\Bar{\boldsymbol{p}}|=\frac{s-m_N^2}{2\sqrt{s}},\quad |\boldsymbol{p}|=\sqrt{\left(\frac{s+m^2_{X}-m_N^2}{2\sqrt{s}}\right)^2-m_{X}^2}.
\end{align}
In the laboratory (lab.) frame of a fixed-target experiment (the rest frame of the initial proton), utilizing $s(E_\gamma)=m_N\left(2 E_\gamma+m_N\right)$ with $E_\gamma$ the photon energy in the lab. frame and Eq.~\eqref{eq.cross section}, we can obtain the $E_\gamma$-dependence of the total cross section $\sigma_{\gamma p\to Xp}(E_\gamma)$. 
The $X(3872)$ parameters are fixed with the binding energy and width being $\delta=40~$keV and $\Gamma_{X}=100~$keV, respectively.

We find that $\bar{D}^{* 0} \Lambda_c(2860)^{+}$ and $\bar{D}^{* 0} \Lambda_c(2940)^{+}$ intermediate states are crucial in determining the line shape of the total cross section. Note that these two channels contribute to different PWs of the $\gamma p \to X(3872)p$ process, i.e., there is no interference. 
The coupling $g_{D^{(*)}p\Lambda_{c1}}$ for the $\Lambda_c(2860)^{+}$ state is numerically identified as the primary source of uncertainties. In practice, the coupling $g_{D^{(*)}p\Lambda_{c2}}$ for the $\Lambda_c(2940)^{+}$ state is fixed to the value given in Table~\ref{tab.coupling}, while the parameter $g_{D^{(*)}p\Lambda_{c1}}$ for the $\Lambda_c(2860)^{+}$ state is varied within an acceptable range of $(7.59,12.23)~\mathrm{GeV}^{-1}$ derived from Appendix.~\ref{app.1}. 

\begin{figure}[tb]
\centering
\includegraphics[width=0.5\textwidth]{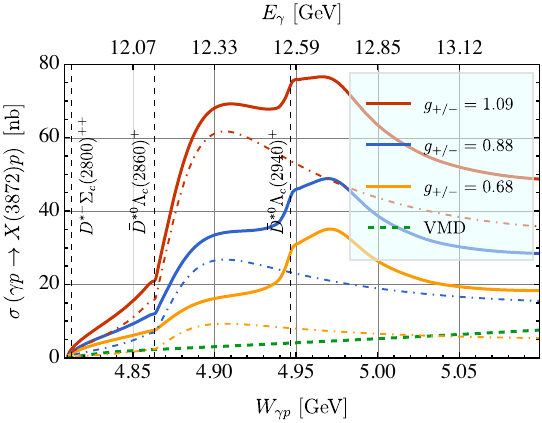} 
\caption{Predicted cross sections for $\gamma p\to X(3872)p$ with parameters in Table~\ref{tab.coupling} taking into account the finite-width effects. The vertical dashed lines indicate the $\bar{D}^{*} \Sigma_c(2800), \bar{D}^{* 0} \Lambda_c(2860)^{+}$ and $\bar{D}^{* 0} \Lambda_c(2940)^{+}$ thresholds. The dot-dashed lines representing only the $\bar{D}^{* 0} \Lambda_c(2860)^{+}$ contribution are also shown for comparison. The prediction of the VMD model is taken from Ref.~\cite{Albaladejo:2020tzt}.} \label{fig.cross-section-v1}
\end{figure}

\begin{figure}[tb]
\centering
\includegraphics[width=.5\textwidth]{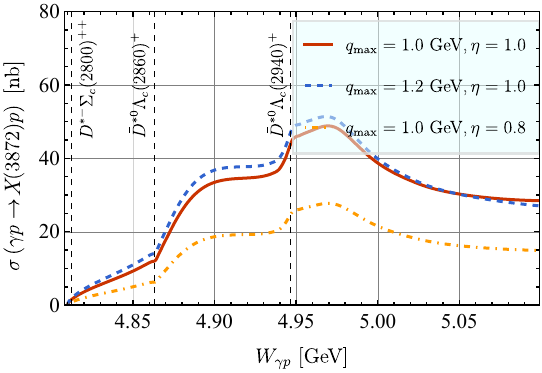} 
\caption{Dependence of the predicted cross section for $\gamma p\to X(3872)p$ on the cutoff parameters $q_\text{max}$ and $\eta$. Here $g_{+/-}$ is fixed to $0.88$. The solid, dashed and dot-dashed curves are obtained by taking $(q_\text{max}=1.0~\mathrm{GeV}, \eta=1.0)$, $(q_\text{max}=1.2~\mathrm{GeV}, \eta=1.0)$ and $(q_\text{max}=1.0~\mathrm{GeV}, \eta=0.8)$, respectively.} \label{fig.cross-section-v3}
\end{figure}

The predicted cross section with the parameters in Table~\ref{tab.coupling} is shown in Fig.~\ref{fig.cross-section-v1}.
We present the line shapes corresponding to the ratio of coupling constants $g_{+/-}\equiv g_{D^{(*)}p\Lambda_{c1}}/g_{D^{(*)}p\Lambda_{c2}}=1.09,0.88$, and $0.68$. In addition, we demonstrate several typical variations in the total cross section corresponding to different parameters of the cutoff~\eqref{eq.cutoff} and the form factor~\eqref{eq.FF} in Fig.~\ref{fig.cross-section-v3}. The $q_\text{max}$-dependence of the cross section is very weak, whereas the $\eta$-dependence exhibits a sizeable effect.
In any case, the total cross section for $\gamma p \rightarrow X(3872) p$ is of $\mathcal{O}(10~\text{nb})$. Moreover, varying the ultraviolet cutoffs does not alter the line shapes which are controlled by infrared singularities.

\begin{figure}[tb]
    \centering
    \includegraphics[width=.4\textwidth]{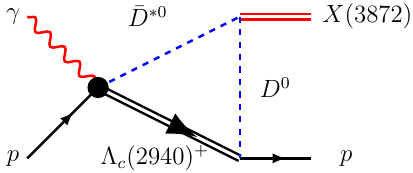} 
    \caption{Shrinking the highly virtual $t$-channel $D^{(*)0}$ propagator attached to the photon in Fig.~\ref{fig.Feyn} to a point (denoted by a filled circle) leads to a triangle diagram.}\label{fig.triangle}
\end{figure}

\begin{figure}[tb]
\centering
\includegraphics[width=.5\textwidth]{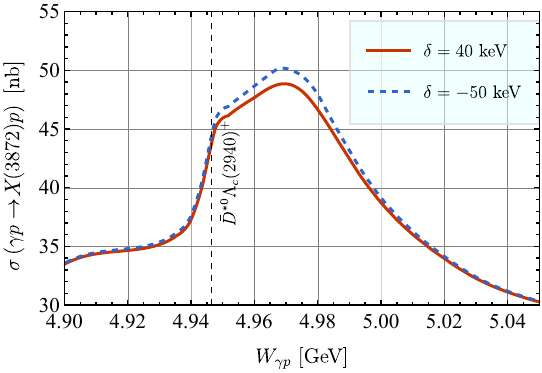} 
\caption{Total cross section for $\gamma p \rightarrow X(3872) p$ as a function of the c.m. energy $W_{\gamma p}$ relative to the $\bar{D}^{* 0}\Lambda_c(2940)^+$ threshold. Here $g_{+/-}$ is fixed to $0.88$. The solid and dashed curves are for the $X(3872)$ binding energy $\delta=40~\mathrm{keV}$ and $-50~\mathrm{keV}$, respectively.} \label{fig.cross-section-v2}
\end{figure}

The considered open-charm coupled-channel mechanism is manifested in the nontrivial structures shown in Fig.~\ref{fig.cross-section-v1}. 
Firstly, there are cusp effects at the open-charm thresholds of the $\bar{D}^{* 0} \Lambda_c(2860)^{+}$ and $\bar{D}^{* 0} \Lambda_c(2940)^{+}$ channels. Secondly, there are also non-trivial enhancements from the three-body triangle singularities (for a recent review, see Ref.~\cite{Guo:2019twa}), which are the subleading Landau singularities of the box diagrams in Fig.~\ref{fig.Feyn} and can be understood through Fig.~\ref{fig.triangle} (here we use $\bar{D}^{* 0} \Lambda_c(2940)^{+}$ channel as an illustration).
The triangle singularity induced line shape is sensitive to the $X(3872)$ binding energy, $\delta$~\cite{Guo:2019qcn, Braaten:2019gfj, Sakai:2020ucu, Ortega:2020lhr, Yan:2022eiy}, if the widths of the intermediate particles are tiny. 
However, once we take into account the finite-width effect of $\Lambda_{ci}$, the sensitivity to the binding energy will be reduced to a level that is hardly visible for the foreseeable experimental accuracy. 
As can be seen from Fig.~\ref{fig.cross-section-v2}, there is a smooth peak due to the finite-width impact of $\Lambda_c(2940)^+$. 
The peak, in particular for negative $\delta$, would become sharper were the $\Lambda_c(2940)^+$ width smaller. 

One observes that the VMD model prediction~\cite{Albaladejo:2020tzt}
is generally smaller than our results obtained with the parameters in Table~\ref{tab.coupling}. Moreover, it is important to note that the line shape derived from the VMD model does not have nontrivial structures without further considering coupled channels. We underscore that the aforementioned comparison provides merely an order-of-magnitude estimate and should not be construed as a precise quantitative prediction. The line shape serves as a more significant criterion for differentiating among various models.

From the above analysis, we conclude that the open-charm loop mechanism advocated here does indeed have the opportunity to make a crucial, possibly dominating contribution to the $X(3872)$ photoproduction off the proton in the near-threshold region, and there could be some broad bumps in the line shape of the cross section due to the open-charm thresholds and triangle singularities.

\section{Summary}\label{sec.summary}

The near-threshold photoproduction of a heavy quarkonium off the proton target is currently of high interest, because it is related to the possibility of measuring the trace anomaly contribution to the proton mass and can be explored to search for hidden-charm pentaquarks.
Thus, it is crucial to understand the mechanism of such processes.
While the VMD model has been widely assumed in the literature, there has been strong evidence that the open-charm coupled-channel mechanism~\cite{Du:2020bqj} may be crucial for the $J/\psi$ photoproduction~\cite{GlueX:2023pev,JointPhysicsAnalysisCenter:2023qgg}.

In this work, we estimate the $X(3872)$ photoproduction off the proton in the near-threshold region considering the open-charm coupled-channel mechanism. We focus on a few intermediate states with nearby thresholds. 
We find that the total cross section for the $X(3872)$ photoproduction in the near-threshold region should be of the order of tens of nanobarns.
In contrast to the VMD model, the coupled-channel mechanism leads to visible structures arising from threshold cusps and triangle singularities due to the $\bar{D}^{*0} \Lambda_c(2860)^+$ and $\bar{D}^{*0} \Lambda_c(2940)^+$ channels. 
Therefore, a measurement of the energy dependence of the cross section is crucial to disentangle the production mechanisms.
The framework presented here is readily applicable to the photoproduction of other charmonium(-like) states, such as $\gamma p \rightarrow Z_c(3900) N$.

The photoproduction of heavy quarkonium(-like) states can be measured at current and forthcoming experiments at facilities such as the 12 GeV Continuous Electron Beam Accelerator Facility at Jefferson Laboratory and its possible 22~GeV upgrade~\cite{Accardi:2023chb}, EIC~\cite{AbdulKhalek:2021gbh}, and EicC~\cite{Anderle:2021wcy}. With these facilities, which are complementary to those of existing experiments producing the $X Y Z$ states, new insights into the nature of charmonium(-like) states and the heavy-quarkonoium--nucleon interactions will be obtained.

\begin{acknowledgements}
XHC would like to thank De-Liang Yao and Mao-Jun Yan for helpful discussions on the treatment of the spin-${3}/{2}$ baryons. This work is supported in part by the National Key R\&D Program of China under Grant No. 2023YFA1606703; by the Chinese Academy of Sciences under Grants No.~XDB34030000 and No. YSBR-101; by the National Natural Science Foundation of China (NSFC) under Grants No. 12125507, No. 12361141819 No. 12047503 and No. 12347120; and by the Postdoctoral Fellowship Program of China Postdoctoral Science Foundation (CPSF) under Grant No. GZC20232773 and the CPSF Grant No. 2023M743601.
\end{acknowledgements}

\begin{appendix}

\section{Determination of the coupling constants}\label{app.1}

\subsection{Magnetic couplings $g_{\gamma D^{(*)0} D^{*0}}$ and $ g_{\gamma D^{+}D^{*-}}$}

Since the $D^*\to D\gamma$ radiative decays are well measured~\cite{ParticleDataGroup:2022pth}, the corresponding couplings can be determined directly from the data. The amplitude of the decay $D^*(k_1)\to D(k_2)\gamma$ can be written as
\begin{align}
    \mathcal{A}\left(D^* \rightarrow D \gamma\right)=-4i g_{\gamma D D^*} \epsilon^{\mu \nu \alpha \beta} \epsilon_\mu \epsilon_\nu^* k_{1 \alpha} k_{2 \beta},
\end{align}
which describes both the neutral and charged channels. The decay width can be obtained as
\begin{align}
    \Gamma=\frac{m_{D^*}^2-m_D^2}{16\pi m_{D^*}^2}\frac{1}{2\times 1+1}\sum_\text{spin}\left|4 g_{\gamma D D^*} \epsilon^{\mu \nu \alpha \beta} \epsilon_\mu \epsilon_\nu^* k_{1 \alpha} k_{2 \beta}\right|^2 = g_{\gamma D D^*}^2\frac{\left(m_{D^*}^2-m_D^2\right)^3}{6 \pi m_{D^*}^2}.\label{eq.Gamma}
\end{align}

Using the value of the total $D^{*0}$ width, $\Gamma_{\text {tot }}\left(D^{* 0}\right) \simeq 55.3~\mathrm{keV}$~\cite{Guo:2019qcn}, we find $\Gamma\left(D^{* 0} \rightarrow D^0 \gamma\right) \simeq 19.5~\mathrm{keV}$, where the branching fraction $\operatorname{Br}\left(D^{* 0} \rightarrow D^0 \gamma\right)=35.3\%$~\cite{ParticleDataGroup:2022pth} was used. Using Eq.~\eqref{eq.Gamma}, we get $\left|g_{\gamma D^0 D^{*0}}\right|=0.142~\mathrm{GeV}^{-1}$. For the radiative decay of the charged $D^{*\pm}$ mesons we have $\Gamma\left(D^{* \pm} \rightarrow D^\pm \gamma\right)\simeq1.6 \% \times 83.4~\mathrm{keV}= 1.33~\mathrm{keV}$~\cite{ParticleDataGroup:2022pth}. Consequently, $\left|g_{\gamma D^+ D^{*-}}\right|=0.035 \mathrm{GeV}^{-1}$.

We further notice that the Lagrangian~\eqref{eq.L1} can be rewritten as a nonrelativistic form:
\begin{align}
    \mathcal{L}\supset -4 g_{\gamma D_a D_b^*} \sqrt{m_{D_a} m_{D_b^*}} v_\alpha \partial_\beta A_\lambda \epsilon^{\lambda \mu \alpha \beta} {D}_{b \mu}^{*} D_a ^\dagger + \text{h.c.},
\end{align}
where $a,b=1,2$ are the light flavor indices. Since the $c$-quark enters both $D$ and $D^*$ meson, the 4-velocity could be introduced to either of them, and the corresponding mass $\left(m_D\right.$ or $\left.m_{D^*}\right)$ would appear. We use a symmetric form $\sqrt{m_D m_{D^*}}$, i.e., $v_\mu \sim i \overleftrightarrow{\partial_\mu}/\sqrt{m_D m_{D^*}}$. Comparing the Lagrangian with Eq.~(18) of Ref.~\cite{Guo:2014taa}, we find
\begin{align}
    g_{\gamma D_a D_b^*}=\frac{e}{4}\left(\beta Q_{a b}+\frac{Q_c}{m_c} \delta_{a b}\right),
\end{align}
where $Q=\operatorname{diag}(2 / 3,-1 / 3)$ is the light-quark charge matrix, $Q_c=2 / 3$ is the charmed quark charge, $m_c$ is the charm quark mass, the term proportional to $Q_c / m_c$ comes from the magnetic moment of the charm quark, and the $\beta$ term is from the nonperturbative light-quark cloud in the charmed mesons. Then it is easy to find 
\begin{align}
    g_{\gamma D^0 D^{*0}}=\frac{e}{6}\left(\beta+\frac{1}{m_c}\right), \quad g_{\gamma D^+ D^{*-}}=-\frac{e}{12}\left(\beta-\frac{2}{m_c}\right).\label{eq.DD*_beta}
\end{align}
The numerical values of the couplings $\left|g_{\gamma D^0 D^{*0}}\right|=0.142~\mathrm{GeV}^{-1}$ and $\left|g_{\gamma D^+ D^{*-}}\right|=0.035 \mathrm{GeV}^{-1}$ can be reconciled with the expressions~\eqref{eq.DD*_beta} for
\begin{align}
    \beta^{-1}=0.428~\mathrm{GeV},\quad m_c=2.104~\mathrm{GeV},\label{eq.beta1}
\end{align}
where
\begin{align}
    g_{\gamma D^0 D^{*0}}=0.142\ \mathrm{GeV}^{-1}, \quad g_{\gamma D^+ D^{*-}}^c=-0.035\ \mathrm{GeV}^{-1}.
\end{align}
They are close to the values obtained using old measurements from Ref.~\cite{Hu:2005gf}:
\begin{align}
    \beta^{-1}=0.379~\mathrm{GeV}, \quad m_c=1.863~\mathrm{GeV}.\label{eq.beta2}
\end{align}

As a by-product, we also get
\begin{align}
    g_{\gamma D_a^* D_b^*}=e m_{D^*}\left(\beta Q_{a b}-\frac{Q_c}{m_c} \delta_{a b}\right),
\end{align}
it is easy to find that
\begin{align}
    g_{\gamma D^{*0} D^{*0}}=\frac{2e}{3} m_{D^*}\left(\beta-\frac{1}{m_c}\right), \quad g_{\gamma D^{*+} D^{*-}}=-\frac{e}{3} m_{D^*}\left(\beta+\frac{2}{m_c}\right),
\end{align}
Based on the parameters~\eqref{eq.beta1}, we find
\begin{align}
    g_{\gamma D^{*0} D^{*0}}=0.755, \quad g_{\gamma D^{*+} D^{*-}}=-0.666,
\end{align}
while Eq.~\eqref{eq.beta2} leads to $g_{\gamma D^{*0} D^{*0}}=0.852$ and $g_{\gamma D^{*+} D^{*-}}=-0.752$.

\subsection{Strong couplings $g_{DN\Sigma_c}$}

We follow the convention: $S_\ell=1+2i\rho T_\ell,\  T(s, \cos \theta)=16 \pi \sum_{\ell=0}^{\infty}(2 \ell+1) P_\ell(\cos \theta) T_\ell(s)$, where $\rho(s)=\sqrt{\lambda(s,m_1^2,m_2^2)}/s$. The couplings of the near-threshold resonances to the meson-baryon components are obtained from the residues of the scattering amplitude. Close to the pole, the PW amplitude can be written in the form
\begin{align}
    T_\ell(s)=\frac{g_ig_j}{s_R-s},
\end{align}
which is the singular part of the Laurent expansion.
Therefore, the residue of Ref.~\cite{Jimenez-Tejero:2009cyn} $g^\prime$ is related to our definition $g$ through $\left|g\right|^2\simeq 2 m_R \frac{2 m_N}{16\pi}\left|g^\prime\right|^2=\frac{m_N m_R}{4\pi}\left|g^\prime\right|^2$, i.e., $\left|g_{\Sigma_c(2800)}\right|=0.961$. Also the residue of Ref.~\cite{Sakai:2020psu} 
$g^\prime$ has a similar expression: $|g|^2=\frac{\mu m_R^2}{2\pi}\left|g^\prime\right|^2$, i.e., $\left|g_{\Sigma_c(2800)}\right|=1.015$.

A cross-check is based on the naive Weinberg  compositeness relation~\cite{Weinberg:1962hj, Weinberg:1965zz} (for more details, see~\cite{Hyodo:2013nka, Guo:2017jvc}), 
\begin{align}
    |g|^2\simeq 2 m_R\frac{m_R\gamma}{2\mu}(1-Z)=\frac{m_R^2 \gamma}{\mu}(1-Z),
\end{align}
where $\gamma$ denotes the binding momentum defined via $\gamma=\sqrt{2 \mu E_B}$ and $\mu=m_1 m_2 /\left(m_1+m_2\right), E_B=m_1+m_2-m_R$. It should be noted that the formula only works in the case where the near-threshold $S$-wave resonance is located below the threshold and is stable or possesses a small width. Assuming $\Sigma_c(2800)$ to be a pure $ND$ hadronic molecule, i.e., $Z=0$, we find $\left|g_{\Sigma_c(2800)}\right|=1.210$, which is close to the above quoted values.

With the isospin phase convention $\left|D^{+}\right\rangle=-\left|D;I=\frac{1}{2}, I_z=\frac{1}{2}\right\rangle$, we have
\begin{align}
    |N D ; 0,0\rangle=\frac{1}{\sqrt{2}}\left(\left|p D^0\right\rangle+\left|n D^{+}\right\rangle\right), \quad |N D ; 1,0\rangle=\frac{1}{\sqrt{2}}\left(\left|p D^0\right\rangle-\left|n D^{+}\right\rangle\right),
\end{align}
thus
\begin{align}
    \left|p D^0\right\rangle=\frac{1}{\sqrt{2}}(|N D ; 0,0\rangle+|N D ; 1,0\rangle),\quad \left|n D^+\right\rangle=\frac{1}{\sqrt{2}}(|N D ; 0,0\rangle-|N D ; 1,0\rangle).
\end{align}
The $ND$ scattering amplitudes in the isospin basis and those in the physical particle basis are related to each other as
\begin{align}
    & T_{N D(I=0)}=\frac{1}{2}\left(T_{p D^0, p D^0}+T_{n D^{+}, n D^{+}}+2 T_{n D^{+}, p D^0}\right), \\
    & T_{N D(I=1)}=\frac{1}{2}\left(T_{p D^0, p D^0}+T_{n D^{+}, n D^{+}}-2 T_{n D^{+}, p D^0}\right).
\end{align}
From Eq.~\eqref{eq.L3}, we find the amplitudes for the $s$-channel exchange of $\Sigma_c(2800)$ as
\begin{align}
    \mathcal{A}_{I=0}=0,\quad \mathcal{A}_{I=1}=-2g^2_{DN\Sigma_c}\frac{\Bar{u}_N\left(\slashed{k}+m_{\Sigma_c}\right)u_N}{s-m_{\Sigma_c}^2},
\end{align}
where $k^2=s$. Since the $\Sigma(2800)$ is a near-threshold state ($m_{\Sigma_c}\simeq m_N+m_D$) that couples to $ND$ in the $S$-wave, we have in the near-threshold region ($s\simeq (m_N+m_D)^2,t\simeq 0, u\simeq (m_N-m_D)^2$),
\begin{align}
    \sigma_{I=1}=&\frac{1}{16\pi s}\frac{1}{2}\sum_{\text{spin}}\left|\mathcal{A}_{I=1}\right|^2\simeq\frac{1}{16\pi s}\frac{64g_{DN\Sigma_c}^4 m_N^2m_{\Sigma_c}^2}{(s-m_{\Sigma_c}^2)^2}=\frac{4g_{DN\Sigma_c}^4 m_N^2m_{\Sigma_c}^2}{\pi s(s-m_{\Sigma_c}^2)^2} \nonumber\\
    \simeq& \frac{16\pi}{(2\times s_{\Sigma_c}+1)s}(2J+1)\left|T_{\ell=0}\right|^2=\frac{16\pi}{\left(2\times \frac{1}{2}+1\right)s}\left(2\times\frac{1}{2}+1\right)\frac{\left|g_{\Sigma_c}\right|^4}{(s-m_{\Sigma_c}^2)^2}.
\end{align}
Finally, we find 
\begin{align}
    \left|g_{DN\Sigma_c}\right|=\sqrt{\frac{2\pi}{m_N m_{\Sigma_c}}}\left|g_{\Sigma_c}\right|=1.57,
\end{align}
where the value  $\left|g_{\Sigma_c}\right|=1.015$~\cite{Sakai:2020psu} is utilized.

\subsection{Strong couplings $g_{Dp\Lambda_{ci}}$}

In the above Lagrangians~\eqref{eq.L5} and~\eqref{eq.L2}, the coupling constants $g_{Dp\Lambda_{ci}}$ can be obtained by fitting the measured partial width of the $\Lambda_c(2860,2940)^{+} \rightarrow D^0 p$ decay~\cite{He:2011jp},
\begin{align}
    \Gamma\left(\Lambda_c(2860,2940)^{+} \rightarrow D^0 p\right)=g_{Dp\Lambda_{ci}}^2 \frac{|\boldsymbol{k}|^3}{12\pi m_{\Lambda_{ci}}}\left(E_N\pm m_N\right),
\end{align}
with $i=1,2$ correspond to $\pm$, respectively. $\boldsymbol{k}$ and $E_N$ are the on-shell three-momentum and the energy of the final proton in the c.m. frame, given by,
\begin{align}
    |\boldsymbol{k}|=\frac{\sqrt{\lambda(m_{\Lambda_{ci}}^2,m_N^2,m_D^2)}}{2 m_{\Lambda_{ci}}},\quad E_N=\frac{m_{\Lambda_{ci}}^2-m_D^2+m_N^2}{2 m_{\Lambda_{ci}}}.
\end{align}
By analyzing the branching ratios $\mathcal{B}\left(\Lambda_c(2860,2940)^{+} \rightarrow D^0 p\right)$, one can determine the coupling constants $g_{Dp\Lambda_{ci}}$. However, experimental efforts, such as those by LHCb~\cite{LHCb:2017jym}, have primarily focused on measuring the total width of $\Lambda_c(2860,2940)^{+}$, without providing the partial decay widths for the $\Lambda_c(2860,2940)^{+} \rightarrow D^0 p$. Consequently, the extraction of $g_{Dp\Lambda_{ci}}$ necessitates the use of theoretical models. Different theoretical frameworks have led to different values for the decay widths of $\Lambda_c(2860,2940)^{+} \rightarrow D^0 p$. 

For the case of the $\Lambda_c(2860)^{+}$, Lin and Liu derived $g_{Dp\Lambda_{c1}}=10.25~\mathrm{GeV}^{-1}$~\cite{Lin:2021wrb} by adopting $\mathcal{B}\left(\Lambda_c(2860)^{+} \rightarrow D^0 p\right)=48\%$~\cite{Chen:2017aqm}, which is from the estimation of the ${}^3 P_0$ model. A smaller value of $41\%$ was obtained with a similar model in Ref.~\cite{Gong:2021jkb}. Therefore, we utilize an averaged value $\mathcal{B}\left(\Lambda_c(2860)^{+} \rightarrow D^0 p\right)\simeq 45\%$ and get $g_{Dp\Lambda_{c1}}=9.91~\mathrm{GeV}^{-1}$. To roughly estimate the effect of the uncertainties of the coupling constant on the cross section, we suppose the total width uncertainty $\Gamma=68^{+12}_{-22}$~MeV and the branching ratio uncertainty $(45\pm 4)\%$ are dominant. Thus the value of $g_{Dp\Lambda_{c1}}$ is about in the range of $(7.59,12.23)~\mathrm{GeV}^{-1}$.

The situation faced by $\Lambda_c(2940)^{+}$ is more subtle due to the uncertainty of spin-parity assignment before 2017. He et al. chose a small value $\mathcal{B}\left(\Lambda_c(2940)^{+} \rightarrow D^0 p\right)<10\%$ and determined $g_{Dp\Lambda_{c2}}=5.26~\mathrm{GeV}^{-1}$~\cite{He:2011jp}. However, in recent years, most models in the literature suggest a high branching ratios, such as: chiral quark model $47\%$~\cite{Ortega:2012cx} and ${}^3 P_0$ model $44\%$~\cite{Lu:2018utx}, $22\%$~\cite{Gong:2021jkb}. We again utilize an averaged value $\mathcal{B}\left(\Lambda_c(2940)^{+} \rightarrow D^0 p\right)\simeq 38\%$ and get $g_{Dp\Lambda_{c2}}=11.25~\mathrm{GeV}^{-1}$.

\section{Rarita-Schwinger vector-spinors}\label{app.2}

For the $\bar{D}^{(*)0}\Lambda_{c}(2860)^+$ and $\bar{D}^{(*)0}\Lambda_{c}(2940)^+$ channels, we have to include the eigen-fields of the $\Lambda_{ci}^+$ baryons, which have the quantum numbers $J^P = \frac{3}{2}^{(\pm)}$. This is usually done in the Rarita-Schwinger framework~\cite{Rarita:1941mf}, which allows for a covariant field-theoretical description of spin-$\frac{3}{2}$ particles. 
The field is represented by a so-called vector-spinor denoted by $\psi^\mu (\mu=0,1,2,3)$, where each $\psi^\mu$ is a Dirac field. Under a proper orthochronous Lorentz transformation $x^{\prime \mu}=\Lambda^\mu_{\ \nu} x^\nu$, the Rarita-Schwinger field has the mixed transformation properties of a four-vector field and a four-component Dirac field, $\psi^{\prime \mu}\left(x^{\prime}\right)=\Lambda_{\ \nu}^\mu S(\Lambda) \psi^\nu(x)$, where $S(\Lambda)$ is the usual matrix representation acting on Dirac spinors. For a relativistic description of a spin-$\frac{3}{2}$ particle, we need $2 \times 4=8$ independent complex fields, where the factor of 2 accounts for the description of particles and antiparticles, and the factor of 4 results from four spin projections in the rest frame ($2\times \frac{3}{2}+1=4$). 
In other words, we need to generate 8 complex conditions among the $4 \times 4=16$ complex fields of the vector-spinor in order to eliminate the additional unphysical degrees of freedom.

The most general free Lagrangian serving that purpose reads~\cite{Moldauer:1956zz} (see Ref.~\cite{Pilling:2004wk} for more details)
\begin{align}
    \mathcal{L}_{\frac{3}{2}}=\bar{\psi}_\mu \Lambda^{\mu \nu}(A) \psi_\nu,
\end{align}
where
\begin{align}
    \Lambda^{\mu \nu}(A)= -\left[\left(i \slashed{\partial}-m\right) g^{\mu \nu}+i A\left(\gamma^\mu \partial^\nu+\gamma^\nu \partial^\mu\right)+\frac{i}{2}\left(3 A^2+2 A+1\right) \gamma^\mu\slashed{\partial} \gamma^\nu+m\left(3 A^2+3 A+1\right) \gamma^\mu \gamma^\nu\right],
\end{align}
with $A \neq-\frac{1}{2}$ an arbitrary real parameter. The Lagrangian introduced by Rarita and Schwinger corresponds to $A=-\frac{1}{3}$. From the Euler-Lagrange equation, we obtain the equation of motion (EOM)
\begin{align}
    \Lambda^{\mu\nu}(A)\psi_\nu=0.
\end{align}
In addition, the $\psi^\mu$ fields satisfy the equations (see for instance Ref.~\cite{Scherer:2012xha})
\begin{align}
    \left(i \slashed\partial-m\right) \psi^\mu&=0, \label{eq.EOM1}\\
    \gamma_\mu \psi^\mu&=0, \label{eq.EOM2}\\
    \partial_\mu \psi^\mu&=0.\label{eq.EOM3}
\end{align}
Each of the Eqs.~\eqref{eq.EOM2} and \eqref{eq.EOM3} generates four complex (subsidiary) conditions. Therefore we end up with the correct number of $16-4-4=8$ independent components. Note that Eq.~\eqref{eq.EOM1} does not reduce the number of independent fields: given the subsidiary conditions, it may rather be interpreted as the EOM.

Additionally, following Ref.~\cite{Huang:2003ym} the proper Rarita-Schwinger vector-spinors (not in the helicity basis) for $s=-\frac{3}{2},-\frac{1}{2},\frac{1}{2},\frac{3}{2}$ can be constructed as
\begin{align}
    u^\mu(p, s) & =\sum_{\lambda=-1,0,1}\sum_{\sigma=-\frac{1}{2},\frac{1}{2}}\left\langle 1, \lambda; \frac{1}{2}, \sigma \right|\left. \frac{3}{2}, s\right\rangle \epsilon^\mu(p, \lambda) u(p, \sigma), \label{eq.RSspinor}\\
    \epsilon^\mu(p, \lambda) & =\left(\frac{\hat{\boldsymbol{\epsilon}}_\lambda \cdot \boldsymbol{p}}{m}, \hat{\boldsymbol{\epsilon}}_\lambda+\frac{\boldsymbol{p}\left(\hat{\boldsymbol{\epsilon}}_\lambda \cdot \boldsymbol{p}\right)}{m\left(E+m\right)}\right), \label{eq.vecpor}\\
    u(p, \sigma) & =\sqrt{E+m}\left(\chi_\sigma, \frac{\boldsymbol{\sigma} \cdot \boldsymbol{p}}{E+m} \chi_\sigma\right)^T,\label{eq.sppor}
\end{align}
where $\left\langle 1, \lambda; \frac{1}{2}, \sigma \right|\left. \frac{3}{2}, s\right\rangle$  denotes the pertinent CG coefficients and
\begin{align}
    &\chi_{1/2}=(1,0),\quad \chi_{-1/2}=(0,1),\\
    &\hat{\boldsymbol{\epsilon}}_0=(0,0,1),\quad \hat{\boldsymbol{\epsilon}}_{+}=-\frac{1}{\sqrt{2}}(1, i, 0), \quad \hat{\boldsymbol{\epsilon}}_{-}=\frac{1}{\sqrt{2}}(1,-i, 0).
\end{align}
A similar construction can be found for the anti-fermion solutions.
The polarization vectors of the vector can be constructed through the Lorentz transformation~\cite{Gulmez:2016scm} and have explicit expressions as Eq.~\eqref{eq.vecpor}:
\begin{align}
    \epsilon^\mu(p,0)&=\left(\frac{|\boldsymbol{p}|}{m} \cos\theta, \frac{1}{2}\left(\frac{E}{m}-1\right) \sin 2 \theta, 0, \frac{E}{m}\cos^2\theta+\sin^2\theta\right),\nonumber\\
    \epsilon^\mu(p,1)&=\left(-\frac{1}{\sqrt{2}} \frac{|\boldsymbol{p}|}{m} \sin \theta, -\frac{1}{\sqrt{2}}\left(\frac{E}{m} \sin ^2 \theta+\cos ^2 \theta\right), -\frac{i}{\sqrt{2}}, -\frac{1}{2 \sqrt{2}}\left(\frac{E}{m}-1\right) \sin 2 \theta\right),\nonumber\\
    \epsilon^\mu(p,-1)&=\left(\frac{1}{\sqrt{2}} \frac{|\boldsymbol{p}|}{m} \sin \theta, \frac{1}{\sqrt{2}}\left(\frac{E}{m} \sin ^2 \theta+\cos ^2 \theta\right), -\frac{i}{\sqrt{2}}, \frac{1}{2 \sqrt{2}}\left(\frac{E}{m}-1\right) \sin 2 \theta\right),
\end{align}
with $p=(E,|\boldsymbol{p}|\sin\theta,0,|\boldsymbol{p}|\cos\theta)$. The Dirac spinors take the general form as Eq.~\eqref{eq.sppor}:
\begin{align}
    u(p,\frac{1}{2})&=\sqrt{E+m}\left(1,0,\frac{|\boldsymbol{p}|\cos\theta}{E+m},\frac{|\boldsymbol{p}|\sin\theta}{E+m}\right)^T,\nonumber\\
    u(p,-\frac{1}{2})&=\sqrt{E+m}\left(0,1,\frac{|\boldsymbol{p}|\sin\theta}{E+m},-\frac{|\boldsymbol{p}|\cos\theta}{E+m}\right)^T.
\end{align}

From Eq.~\eqref{eq.RSspinor}, the Rarita-Schwinger vector-spinors can be constructed as
\begin{align}
    u^\mu\left(p,\frac{3}{2}\right)&=\left\langle 1, 1; \frac{1}{2}, \frac{1}{2} \right|\left. \frac{3}{2}, \frac{3}{2}\right\rangle\epsilon^\mu(p,1)u(p,\frac{1}{2})=\epsilon^\mu(p,1)u(p,\frac{1}{2}),\nonumber\\
    u^\mu\left(p,\frac{1}{2}\right)&=\left\langle 1, 1; \frac{1}{2}, -\frac{1}{2} \right|\left. \frac{3}{2}, \frac{1}{2}\right\rangle\epsilon^\mu(p,1)u(p,-\frac{1}{2})+\left\langle 1, 0; \frac{1}{2}, \frac{1}{2} \right|\left. \frac{3}{2}, \frac{1}{2}\right\rangle\epsilon^\mu(p,0)u(p,\frac{1}{2}) \nonumber\\
    &=\frac{1}{\sqrt{3}}\epsilon^\mu(p,1)u(p,-\frac{1}{2})+\sqrt{\frac{2}{3}}\epsilon^\mu(p,0)u(p,\frac{1}{2}),\nonumber\\
    u^\mu\left(p,-\frac{1}{2}\right)&=\left\langle 1, -1; \frac{1}{2}, \frac{1}{2} \right|\left. \frac{3}{2}, -\frac{1}{2}\right\rangle\epsilon^\mu(p,-1)u(p,\frac{1}{2})+\left\langle 1, 0; \frac{1}{2}, -\frac{1}{2} \right|\left. \frac{3}{2}, -\frac{1}{2}\right\rangle\epsilon^\mu(p,0)u(p,-\frac{1}{2}) \nonumber\\
    &=\frac{1}{\sqrt{3}}\epsilon^\mu(p,-1)u(p,\frac{1}{2})+\sqrt{\frac{2}{3}}\epsilon^\mu(p,0)u(p,-\frac{1}{2}),\nonumber\\
    u^\mu\left(p,-\frac{3}{2}\right)&=\left\langle 1, -1; \frac{1}{2}, -\frac{1}{2} \right|\left. \frac{3}{2}, -\frac{3}{2}\right\rangle\epsilon^\mu(p,-1)u(p,-\frac{1}{2})=\epsilon^\mu(p,-1)u(p,-\frac{1}{2}).\label{eq.RSspinor_2}
\end{align}
In the rest frame, Eqs.~\eqref{eq.RSspinor_2} are given by
\begin{align}
    u^\mu(p, 3 / 2) & =\sqrt{2m}\{(0,0,0,0),(-1 / \sqrt{2}, 0,0,0),(-i / \sqrt{2}, 0,0,0),(0,0,0,0)\}^T, \nonumber\\
    u^\mu(p, 1 / 2) & =\sqrt{2m}\{(0,0,0,0),(0,-1 / \sqrt{6}, 0,0),(0,-i / \sqrt{6}, 0,0),(\sqrt{2 / 3}, 0,0,0)\}^T, \nonumber\\
    u^\mu(p, -1 / 2) & =\sqrt{2m}\{(0,0,0,0),(1 / \sqrt{6}, 0,0,0),(-i / \sqrt{6}, 0,0,0),(0, \sqrt{2 / 3}, 0,0)\}^T, \nonumber\\
    u^\mu(p, -3 / 2) & =\sqrt{2m}\{(0,0,0,0),(0,1 / \sqrt{2}, 0,0),(0,-i / \sqrt{2}, 0,0),(0,0,0,0)\}^T,\label{eq.RS1}
\end{align}
with $p^\mu=(m,0,0,0)$. In a frame of arbitrary momentum, e.g., $p^\mu=(E,|\boldsymbol{p}|\sin\theta,0,|\boldsymbol{p}|\cos\theta)$, Eqs.~\eqref{eq.RSspinor_2} have expressions:
\begin{align}
    u^\mu(p, 3 / 2)=\sqrt{2m}\Biggr\{ & \left(-\frac{|\boldsymbol{p}| \sin \theta}{2 m} \sqrt{\frac{E+m}{m}}, 0,-\frac{|\boldsymbol{p}|^2 \sin 2 \theta}{4 m^2} \sqrt{\frac{m}{E+m}},-\frac{|\boldsymbol{p}|^2 \sin ^2 \theta}{2 m^2} \sqrt{\frac{m}{E+m}}\right), \nonumber\\
    & \left(-\frac{1}{2 m}\left(E \sin ^2 \theta+m \cos ^2 \theta\right) \sqrt{\frac{E+m}{m}}, 0,\right. \nonumber\\
    &\quad-\frac{|\boldsymbol{p}| \cos \theta}{2 m^2}\left(E \sin ^2 \theta+m \cos ^2 \theta\right) \sqrt{\frac{m}{E+m}}, \nonumber\\
    & \left.\quad-\frac{|\boldsymbol{p}| \sin \theta}{2 m^2}\left(E \sin ^2 \theta+m \cos ^2 \theta\right) \sqrt{\frac{m}{E+m}}\right), \nonumber\\
    & \left(-\frac{i}{2} \sqrt{\frac{E+m}{m}}, 0,-\frac{i|\boldsymbol{p}| \cos \theta}{2 m} \sqrt{\frac{m}{E+m}},-\frac{i|\boldsymbol{p}| \sin \theta}{2 m} \sqrt{\frac{m}{E+m}}\right), \nonumber\\
    & \left(\frac{\left(m-E\right) \sin 2 \theta}{4 m} \sqrt{\frac{E+m}{m}}, 0, \frac{|\boldsymbol{p}|\left(m-E\right) \cos ^2 \theta \sin \theta}{2 m^2} \sqrt{\frac{m}{E+m}},\right. \nonumber\\
    & \left.\left.\quad\frac{|\boldsymbol{p}|\left(m-E\right) \cos \theta \sin ^2 \theta}{2 m^2} \sqrt{\frac{m}{E+m}}\right)\right\}^T,\nonumber\\
    u^\mu(p, 1 / 2)=\sqrt{2m}\Biggr\{ & \left(\frac{|\boldsymbol{p}| \cos \theta}{\sqrt{3} m} \sqrt{\frac{E+m}{m}},-\frac{|\boldsymbol{p}| \sin \theta}{2 \sqrt{3} m} \sqrt{\frac{E+m}{m}},\right. \nonumber\\
    & \left.\quad\frac{|\boldsymbol{p}|^2(1+3 \cos 2 \theta)}{4 \sqrt{3} m^2} \sqrt{\frac{m}{E+m}}, \frac{\sqrt{3}|\boldsymbol{p}|^2 \sin 2 \theta}{4 m^2} \sqrt{\frac{m}{E+m}}\right), \nonumber\\
    & \left(\frac{\left(E-m\right) \cos \theta \sin \theta}{\sqrt{3} m} \sqrt{\frac{E+m}{m}},-\frac{\left(E \sin ^2 \theta+m \cos ^2 \theta\right)}{2 \sqrt{3} m} \sqrt{\frac{E+m}{m},}\right. \nonumber\\
    & \quad\frac{|\boldsymbol{p}|\left(E(1+3 \cos 2 \theta)-6 m \cos ^2 \theta\right) \sin \theta}{4 \sqrt{3} m^2} \sqrt{\frac{m}{E+m}}, \nonumber\\
    & \left.\quad\frac{|\boldsymbol{p}|\left(3 E \sin ^2 \theta+m\left(\cos ^2 \theta-2 \sin ^2 \theta\right)\right) \cos \theta}{2 \sqrt{3} m^2} \sqrt{\frac{m}{E+m}}\right), \nonumber\\
    & \left(0,-\frac{i}{2 \sqrt{3}} \sqrt{\frac{E+m}{m}},-\frac{i|\boldsymbol{p}| \sin \theta}{2 \sqrt{3} m} \sqrt{\frac{m}{E+m}}, \frac{i|\boldsymbol{p}| \cos \theta}{2 \sqrt{3} m} \sqrt{\frac{m}{E+m}}\right), \nonumber\\
    & \left(\frac{\left(m \sin ^2 \theta+E \cos ^2 \theta\right)}{\sqrt{3} m} \sqrt{\frac{E+m}{m}}, \frac{\left(m-E\right) \sin 2 \theta}{4 \sqrt{3} m} \sqrt{\frac{E+m}{m}},\right. \nonumber\\
    & \quad\frac{|\boldsymbol{p}|\left(E(1+3 \cos 2 \theta)-6 m \sin ^2 \theta\right) \cos \theta}{4 \sqrt{3} m^2} \sqrt{\frac{m}{E+m}}, \nonumber\\
    & \left.\left.\quad\frac{|\boldsymbol{p}|\left(m\left(2 \sin ^2 \theta-\cos ^2 \theta\right)+3 E \cos ^2 \theta\right) \sin \theta}{2 \sqrt{3} m^2} \sqrt{\frac{m}{E+m}}\right)\right\}^T,\nonumber\\
    u^\mu(p, -1 / 2)=\sqrt{2m}\Biggr\{ & \left(\frac{|\boldsymbol{p}| \sin \theta}{2 \sqrt{3} m} \sqrt{\frac{E+m}{m}}, \frac{|\boldsymbol{p}| \cos \theta}{\sqrt{3} m} \sqrt{\frac{E+m}{m}},\right. \nonumber\\
    & \left.\quad\frac{\sqrt{3}|\boldsymbol{p}|^2 \sin 2 \theta}{4 m^2} \sqrt{\frac{m}{E+m}},-\frac{|\boldsymbol{p}|^2(1+3 \cos 2 \theta)}{4 \sqrt{3} m^2} \sqrt{\frac{m}{E+m}}\right), \nonumber\\
    & \left(\frac{\left(E \sin ^2 \theta+m \cos ^2 \theta\right)}{2 \sqrt{3} m} \sqrt{\frac{E+m}{m}}, \frac{\left(E-m\right) \cos \theta \sin \theta}{\sqrt{3} m} \sqrt{\frac{E+m}{m}},\right. \nonumber\\
    & \quad \frac{|\boldsymbol{p}|\left(3 E \sin ^2 \theta+m\left(\cos ^2 \theta-2 \sin ^2 \theta\right)\right) \cos \theta}{2 \sqrt{3} m^2} \sqrt{\frac{m}{E+m}}, \nonumber\\
    & \left.\quad-\frac{|\boldsymbol{p}|\left(E(1+3 \cos 2 \theta)-6 m \cos ^2 \theta\right) \sin \theta}{4 \sqrt{3} m^2} \sqrt{\frac{m}{E+m}}\right), \nonumber\\
    & \left(-\frac{i}{2 \sqrt{3}} \sqrt{\frac{E+m}{m}}, 0,-\frac{i|\boldsymbol{p}| \cos \theta}{2 \sqrt{3} m} \sqrt{\frac{m}{E+m}},-\frac{i|\boldsymbol{p}| \sin \theta}{2 \sqrt{3} m} \sqrt{\frac{m}{E+m}}\right), \nonumber\\
    & \left(\frac{\left(E-m\right) \sin 2 \theta}{4 \sqrt{3} m} \sqrt{\frac{E+m}{m}}, \frac{\left(m \sin ^2 \theta+E \cos ^2 \theta\right)}{\sqrt{3} m} \sqrt{\frac{E+m}{m}},\right. \nonumber\\
    & \quad \frac{|\boldsymbol{p}|\left(m\left(2 \sin ^2 \theta-\cos ^2 \theta\right)+3 E \cos ^2 \theta\right) \sin \theta}{2 \sqrt{3} m^2} \sqrt{\frac{m}{E+m}}, \nonumber\\
    & \left.\left.\quad-\frac{|\boldsymbol{p}|\left(E(1+3 \cos 2 \theta)-6 m \sin ^2 \theta\right) \cos \theta}{4 \sqrt{3} m^2} \sqrt{\frac{m}{E+m}}\right)\right\}^T,\nonumber\\
    u^\mu(p, -3 / 2)=\sqrt{2m}\Biggr\{ & \left(0, \frac{|\boldsymbol{p}| \sin \theta}{2 m} \sqrt{\frac{E+m}{m}}, \frac{|\boldsymbol{p}|^2 \sin ^2 \theta}{2 m^2} \sqrt{\frac{m}{E+m}},-\frac{|\boldsymbol{p}|^2 \sin 2 \theta}{4 m^2} \sqrt{\frac{m}{E+m}}\right), \nonumber\\
    & \left(0, \frac{1}{2 m}\left(E \sin ^2 \theta+m \cos ^2 \theta\right) \sqrt{\frac{E+m}{m}},\right. \nonumber\\
    & \quad\frac{|\boldsymbol{p}| \sin \theta}{2 m^2}\left(E \sin ^2 \theta+m \cos ^2 \theta\right) \sqrt{\frac{m}{E+m}}, \nonumber\\
    & \left.\quad-\frac{|\boldsymbol{p}| \cos \theta}{2 m^2}\left(E \sin ^2 \theta+m \cos ^2 \theta\right) \sqrt{\frac{m}{E+m}}\right), \nonumber\\
    & \left(0,-\frac{i}{2} \sqrt{\frac{E+m}{m}},-\frac{i|\boldsymbol{p}| \sin \theta}{2 m} \sqrt{\frac{m}{E+m}}, \frac{i|\boldsymbol{p}| \cos \theta}{2 m} \sqrt{\frac{m}{E+m}}\right), \nonumber\\
    & \left(0, \frac{\left(E-m\right) \cos \theta \sin \theta}{2 m} \sqrt{\frac{E+m}{m}}, \frac{|\boldsymbol{p}|\left(E-m\right) \cos \theta \sin ^2 \theta}{2 m^2} \sqrt{\frac{m}{E+m}},\right. \nonumber\\
    & \left.\left.\quad\frac{|\boldsymbol{p}|\left(m-E\right) \cos ^2 \theta \sin \theta}{2 m^2} \sqrt{\frac{m}{E+m}}\right)\right\}^T.\label{eq.RS2}
\end{align}
From Eqs.~\eqref{eq.RS1} and \eqref{eq.RS2} we can derive the spin-energy projection operator,
\begin{align}
    \mathcal{P}^{\mu \nu}_{3/2}(p) =\sum_{\lambda} u^\mu(p, \lambda) \bar{u}^\nu(p, \lambda)=(\slashed{p}+m)\left(-g^{\mu \nu}+\frac{1}{3} \gamma^\mu \gamma^\nu+\frac{2}{3 m^2} p^\mu p^\nu+\frac{p^\nu \gamma^\mu-p^\mu \gamma^\nu}{3 m}\right).
\end{align}

Another subtlety in incorporating spin-$\frac{3}{2}$ particles into the theory is to ensure the correct decoupling of the unphysical spin-$\frac{1}{2}$ components of the field. In the free case, these components are projected out in the resulting EOM, but in the case of interacting spin-$\frac{3}{2}$ fields, the task is more subtle. A first type of effective interaction at low energies was proposed (which is translated to the $Dp \Lambda_c(2940)^+$
case as an example)~\cite{Nath:1971wp}
\begin{align}\label{eq.L1_3/2}
    \mathcal{L}_1=g m_{\Lambda_{c2}} \bar{\Lambda}_\mu\left(g^{\mu \nu}+a \gamma^\mu \gamma^\nu\right) \gamma_5 p \partial_\nu D^0+ \text{h.c.},
\end{align}
where $a$ is an off-shell parameter that is relevant only for loop computations. In tree-level perturbative calculations, it is justified to assign a value of 0 to $a$. Then the Lagrangian~\eqref{eq.L1_3/2} is equivalent to Eq.~\eqref{eq.L2} we used.

This interaction with the smallest number of derivatives is simple. But, as discussed in Ref.~\cite{Pascalutsa:1998pw}, the interacting theory involves not only the physical spin-$\frac{3}{2}$ components of the Rarita-Schwinger field, but also the unphysical spin-$\frac{1}{2}$ components, leading to problems of causality and to a significant contribution from the spin-$\frac{1}{2}$ background underneath the $\Lambda_{c2}^+$ resonance. In Ref.~\cite{Pascalutsa:1998pw}, Pascalutsa showed that the problem can be resolved by requiring that all interactions have the same type of gauge invariance as the kinetic term of the spin-$\frac{3}{2}$ field, i.e. the free massless Lagrangian are invariant under the local (gauge) transformation of the spin-$\frac{3}{2}$ field: $\Lambda_\mu(x) \to \Lambda_\mu(x)+\partial_\mu \varepsilon(x)$, where $\varepsilon(x)$ is an arbitrary spinor. Thus an alternative interaction has been proposed~\cite{Pascalutsa:1998pw}
\begin{align}\label{eq.L2_3/2}
    \mathcal{L}_2=g \epsilon^{\mu \nu \alpha \beta}\left(\partial_\mu \bar{\Lambda}_\nu\right) \gamma_\alpha p \partial_\beta D^{0}+\text{h.c.}.
\end{align}

However, this additional demand for gauge invariance is not necessary in the context of the commonly adapted effective field theory.
In Ref.~\cite{Pascalutsa:2000kd} it was observed that every gauge noninvariant linear coupling of the spin-$\frac{3}{2}$ field can be transformed into a gauge invariant form by choosing a suitable field redefinition. Then, Krebs, Epelbaum and Mei\ss{}ner~\cite{Krebs:2008zb} provided a similar statement for bilinear couplings of the spin-$\frac{3}{2}$ field. Because of the proven equivalence of the two approaches related by a nonlinear field redefinition, all of these considerations imply that the elements of the $S$-matrix, as derived from the two effective Lagrangians, are fundamentally equivalent. Recent phenomenological analysis~\cite{Descotes-Genon:2019dbw} also provides a numerical confirmation of this equivalence from another point of view.

\end{appendix}

\bibliography{ref}
\end{document}